\documentclass[aps,prb,10pt,
superscriptaddress, twocolumn]{revtex4-2}

\usepackage{tikz,pgf}
\usepackage{amsfonts}
\usepackage{amsmath}
\usepackage{amssymb}
\usepackage{mathtools}
\usepackage{braket}
\usepackage[colorlinks=true,linkcolor=blue,citecolor=blue,urlcolor=blue]{hyperref}
\usepackage{standalone}
\usepackage{overpic}
\usepackage[right]{lineno}
\usepackage{algorithmicx,algpseudocode}
\usepackage[linesnumbered,ruled,vlined]{algorithm2e}
\usepackage{dcolumn}
\usepackage{listings}
\usepackage{subcaption}
\usepackage{caption}
\usepackage{enumitem}
\usepackage{bbm}
\usepackage{orcidlink}
\usepackage{graphicx}
\usepackage{xcolor}
\usepackage[T1]{fontenc}

\newcommand{\C}{\mathbb{C}}

\newcommand{\tikzpath}{}
\newcommand{\figpath}{}

\usetikzlibrary{arrows.meta,
                positioning,
                calc,
                shapes.geometric,
                math,
                3d}

\def\lw{1pt}
\def\mpsvirtdist{1.25cm}
\def\mpsoutervirtdist{0.8cm}
\def\mpsinnerphysdist{1.5cm}
\def\mpsouterphysdist{0.8cm}
\def\ttnvirtdist{1.25}
\def\ttnoutervirtdist{1.5}
\def\ttninnerphysdist{1.5}
\def\ttnouterphysdist{0.8}
\def\ttnparalleloutervirt{1}
\def\indexsize{\tiny}
\def\invsqrttwo{0.707}

\makeatletter
\newcommand\thefontsize[1]{{#1 The current font size is: \f@size pt\par}}
\makeatother

\definecolor{mblue}  {rgb}{0.368417, 0.506779, 0.709798}
\definecolor{morange}{rgb}{0.880722, 0.611041, 0.142051}
\definecolor{mgreen} {rgb}{0.560181, 0.691569, 0.194885}
\definecolor{mred}   {rgb}{0.922526, 0.385626, 0.209179}
\definecolor{mpurple}{rgb}{0.647624, 0.37816,  0.614037}
\definecolor{mcyan}  {rgb}{0.363898, 0.618501, 0.782349}
\definecolor{myellow}{rgb}{1, 0.75, 0}

\tikzset{
    statenode/.style={
        circle,
        draw,
        minimum size = 0.6cm,
        fill = mblue!70,
        line width = \lw,
        inner sep=0.01
    },
    matrix/.style={
        diamond,
        draw,
        minimum size = 0.6cm,
        fill = mblue!70,
        line width = \lw,
        inner sep=0.01
    },
    operatornode/.style={
        draw,
        line width=\lw,
        fill=mgreen!40,
        inner sep=0.01,
        minimum size=0.6cm
    },
    submps/.style={
        draw,
        ellipse,
        fill=mcyan,
        line width=\lw,
        minimum width=6mm,
        minimum height=\mpsinnerphysdist*1.2
    },
    tensorcut/.style={
        dashed,
        line width=\lw,
        color=mred
    },
    isonode/.style={
        kite,
        draw,
        minimum size = 0.6cm,
        fill = mblue!40,
        rotate = 90,
        line width = \lw,
        inner sep=0.01
    },
    submpsright/.style={
        draw,
        ellipse,
        fill=mred!40,
        line width=\lw,
        minimum width=6mm,
        minimum height=\mpsinnerphysdist*1.2
    },
    intertnarrow/.style={
        line width=\lw,
        -stealth
    },
    subttn/.style={
        draw,
        ellipse,
        fill=mcyan,
        line width=\lw,
        minimum width=6mm,
        minimum height=\ttninnerphysdist*1.2cm
    },
    subttnright/.style={
        draw,
        ellipse,
        fill=mred!40,
        line width=\lw,
        minimum width=6mm,
        minimum height=\ttninnerphysdist*1.2cm
    },
    csimnode/.style={
        circle,
        draw,
        minimum size = 0.4cm,
        fill = mblue!40,
        line width = \lw,
        inner sep=0.01
    },
    copy/.style={
        circle,
        draw,
        fill=black,
        minimum size=0.15cm,
        inner sep=0.01
    },
}

\newcommand{\setcoords}{
    \pgfsetxvec{\pgfpoint{1cm}{0cm}}
    \pgfsetyvec{\pgfpoint{0.4cm}{0.3cm}}
    \pgfsetzvec{\pgfpoint{0cm}{1cm}}
}
\def\equationautorefname~#1\null{Eq.~(#1)\null}
\def\figureautorefname~#1\null{Fig.~#1\null}
\def\subfigureautorefname~#1\null{Fig.~#1\null}
\def\sectionautorefname~#1\null{Sec.~#1\null}
\def\tableautorefname~#1\null{Table~#1\null}

\begin{document}

\title{Efficient Application of Tensor Network Operators to Tensor Network States Through Successive Deterministic Compression}

\author{Richard M.~Milbradt \orcidlink{0000-0001-8630-9356}}
\email[Corresponding Author:]{r.milbradt@tum.de}
\affiliation{Technical University of Munich, CIT, Department of Computer Science, Boltzmannstra{\ss}e 3, 85748 Garching, Germany}
\author{Shuo Sun~\orcidlink{0009-0006-5775-9730}}
\email{shuo.sun@in.tum.de}
\affiliation{Technical University of Munich, CIT, Department of Computer Science, Boltzmannstra{\ss}e 3, 85748 Garching, Germany}
\author{Christian B.~Mendl~\orcidlink{0000-0002-6386-0230}}
\email{christian.mendl@tum.de}
\affiliation{Technical University of Munich, CIT, Department of Computer Science, Boltzmannstra{\ss}e 3, 85748 Garching, Germany}
\affiliation{Technical University of Munich, Institute for Advanced Study, Lichtenbergstra{\ss}e 2a, 85748 Garching, Germany}
\author{Johnnie Gray~\orcidlink{0000-0001-9461-3024}}
\affiliation{Division of Chemistry and Chemical Engineering, California Institute of Technology, Pasadena, CA 91125, USA}
\author{Garnet Kin-Lic~Chan~\orcidlink{0000-0001-8009-6038}}
\affiliation{Division of Chemistry and Chemical Engineering, California Institute of Technology, Pasadena, CA 91125, USA}
\affiliation{Marcus Center for Theoretical Chemistry, California Institute of Technology, Pasadena, CA 91125, USA}
\email{gkc1000@gmail.com}
\date{\today}

\begin{abstract}
We introduce an algorithm to apply tree tensor network operators to tree tensor network states followed by compression. Especially when applying a matrix product operator to matrix product states, this serves as a key subroutine in tensor network algorithms. We name the method successive deterministic compression (SDC). We compare SDC to existing operator application and compression algorithms in benchmarks involving random tensor network operators and states, as well as circuit simulation with standard entangling gates. We find that successive deterministic compression is consistently among the best-performing techniques when the tensor network state (after the operator application) is not highly compressible. Our results also highlight the flexibility of general tree tensor network structures for simulating circuits with long-range entanglement. 
\end{abstract}

\maketitle

\section{Introduction}\label{sec:introduction}
Tensor networks have proven to be powerful tools for simulating large quantum systems, including quantum circuit simulation, quantum chemistry, and many-body physics. Here, we will focus on one specific subroutine for tensor network methods, which is the efficient evaluation of
\begin{equation}\label{eq:abstr_application}
    \hat{O} \ket{\psi},
\end{equation}
where $\hat{O}$ is a tensor network operator and $\ket{\psi}$ is a tensor network state, followed by compression of $\hat{O} \ket{\psi}$ to a smaller tensor network bond dimension. This subroutine is used in many tensor network simulations, including the computation of ground states and other low-lying states \cite{Rakhuba2016, Baiardi2019, Wang2025, Wang2025B, Dektor2025}, the approximate contraction of two-dimensional tensor networks \cite{Verstraete2004, Lubasch2014}, the time evolution of quantum systems \cite{Zaletel2015, Paeckel2019, Gaggioli2025, Dargel2012, Wall2012, Yang2020}, and the simulation of quantum circuits \cite{Chen2023, Gelss2023}. We will focus particularly on loop-free or tree tensor networks (TTN~\cite{Shi2006, Silvi2019}), which have proven useful in many applications \cite{Shi2006, Larsson2019, Larsson2024, Larsson2025, Li2012, Nakatani2013, Sulz2024, Seitz2023, Dubey2025, Murg2010, Ke2023, Ke2025, Ke2025B, Schuhmacher2025, Sun2025} and which includes the special and most widely used case of linear TTN, known as tensor trains or matrix product states (MPS~\cite{Schollwock2011, Oseledets2011}).

There are already many algorithms for applying tensor network operators with compression. These have various tradeoffs between performance and accuracy, although it is not always simple to analyze their relative performance, which depends on many characteristics of $\hat{O}$ and $\psi$. One recently introduced algorithm is the successive randomised compression, which has been studied in the matrix product operator and matrix product state setting~\cite{Camano2025}, where it has been argued to offer both improved speed and performance. Here, we introduce a method which we name successive deterministic compression (SDC) because it can be seen as a deterministic analog of successive randomised compression (although we will primarily motivate and explain it without this connection). Through asymptotic analysis and numerical trials with random operators and states, as well as quantum circuit simulations, we survey the performance of existing operator application and compression algorithms in  both the matrix product and general tree tensor network settings. This leads us to a deeper understanding of the relative strengths of different methods and establishes the utility of the successive deterministic compression approach. 

\section{Methods}
\subsection{Tree Tensor Networks}\label{sec:TTN}
This section introduces the nomenclature used in this work regarding tree tensor networks. It is assumed that the reader is familiar with the basics of tensor networks, including contractions and decompositions. More pedagogical and in-depth introductions can be found in \cite{Schollwock2011, Bridgeman2017, Silvi2019, Evenbly2022}. For a quick overview, consider the recent reviews \cite{Orus2019, Cirac2021, Banuls2023}.

\subsubsection{Tree Tensor Network States}
\begin{figure*}
    \begin{subfigure}{0.49\textwidth}
        \begin{tikzpicture}
    \useasboundingbox (-1.2*\mpsoutervirtdist,-1.2*\mpsouterphysdist) rectangle (4*\mpsvirtdist+1.2*\mpsouterphysdist,1.2*\mpsouterphysdist);

    \foreach \i/\l in {0/1,1/2,3/L-1,4/L}{
        \node[statenode] (S\i) at (\i*\mpsvirtdist,0) {\indexsize{$\l$}};
        \draw[line width=\lw] (S\i) -- ++(0,\mpsouterphysdist) node[anchor=west] {\indexsize{$\sigma_{\l}$}};
    }
    \draw[line width=\lw, dashed] (-\mpsoutervirtdist,0) -- (S0)node[midway, above] {\indexsize{$\alpha_0$}};
    \draw[line width=\lw, dashed] (S4) -- ++(\mpsoutervirtdist,0)node[midway, above] {\indexsize{$\alpha_L$}};
    \draw[line width=\lw] (S0) -- (S1)node[midway, above] {\indexsize{$\alpha_1$}} -- ++(\mpsoutervirtdist,0)node[midway, above] {\indexsize{$\alpha_2$}};
    \draw[line width=\lw] (S4) -- (S3)node[midway, above] {\indexsize{$\alpha_{L-1}$}} -- ++(-1*\mpsoutervirtdist,0)node[midway, anchor=-80] {\indexsize{$\alpha_{L-2}$}};
    \draw[line width=\lw, dotted] ($(S1) +(\mpsoutervirtdist,0)$) -- ($(S3) + (-1*\mpsoutervirtdist,0)$);
\end{tikzpicture}
        \caption{MPS}
        \label{fig:mps}
    \end{subfigure}
    \begin{subfigure}{0.49\textwidth}
        \begin{tikzpicture}
    \useasboundingbox (-1.2*\mpsoutervirtdist,-1.2*\mpsouterphysdist) rectangle (4*\mpsvirtdist+1.2*\mpsouterphysdist,1.2*\mpsouterphysdist);

    \foreach \i/\l in {0/1,1/2,3/L-1,4/L}{
        \node[operatornode] (O\i) at (\i*\mpsvirtdist,0) {\indexsize{$\l$}};
        \draw[line width=\lw] (O\i) -- ++(0,\mpsouterphysdist);
        \draw[line width=\lw] (O\i) -- ++(0,-1*\mpsouterphysdist);
    }
    \draw[line width=\lw, dashed] (-\mpsoutervirtdist,0) -- (O0);
    \draw[line width=\lw, dashed] (O4) -- ++(\mpsoutervirtdist,0);
    \draw[line width=\lw] (O0) -- (O1) -- ++(\mpsoutervirtdist,0);
    \draw[line width=\lw] (O4) -- (O3) -- ++(-1*\mpsoutervirtdist,0);
    \draw[line width=\lw, dotted] ($(O1) +(\mpsoutervirtdist,0)$) -- ($(O3) + (-1*\mpsoutervirtdist,0)$);
\end{tikzpicture}
        \caption{MPO}
        \label{fig:mpo}
    \end{subfigure}
    \begin{subfigure}{0.49\textwidth}
        \begin{tikzpicture}
    \useasboundingbox (-3.5*\ttnvirtdist,-2.5*\ttnvirtdist) rectangle (2.5*\ttnvirtdist,2.5*\ttnvirtdist);

    \node[statenode] (R) at (0,0){\indexsize{$1$}};
    \node[statenode] (V1) at (\invsqrttwo*\ttnvirtdist,\invsqrttwo*\ttnvirtdist) {\indexsize{$2$}};
    \node[statenode] (P2) at ($(V1) + (0,\ttnvirtdist)$) {\indexsize{$3$}};
    \node[statenode] (P3) at ($(V1) + (\ttnvirtdist,0)$) {\indexsize{$4$}};
    \node[statenode] (V2) at (\invsqrttwo*\ttnvirtdist,-1*\invsqrttwo*\ttnvirtdist) {\indexsize{$5$}};
    \node[statenode] (P4) at ($(V2) + (\ttnvirtdist,0)$) {\indexsize{$6$}};
    \node[statenode] (P5) at ($(V2) + (0,-1*\ttnvirtdist)$) {\indexsize{$7$}};
    \node[statenode] (P1) at ($(R) + (-1*\ttnvirtdist,0)$) {\indexsize{$8$}};
    \node[statenode] (V3) at ($(P1) + (-1*\ttnvirtdist,0)$) {\indexsize{$9$}};
    \node[statenode] (P6) at ($(V3) + (-1*\invsqrttwo*\ttnvirtdist,\invsqrttwo*\ttnvirtdist)$) {\indexsize{$10$}};
    \node[statenode] (P7) at ($(V3) + (-1*\invsqrttwo*\ttnvirtdist,-1*\invsqrttwo*\ttnvirtdist)$) {\indexsize{$11$}};
    \draw[line width=\lw] (P6) -- (V3)node[midway,anchor=south west, inner sep=0.1]{\indexsize{$\alpha_{10}$}} --(P7)node[midway,anchor=north west, inner sep=0.1]{\indexsize{$\alpha_{11}$}};
    \draw[line width=\lw] (V3) -- (P1)node[midway,anchor=south, inner sep=1]{\indexsize{$\alpha_9$}} -- (R)node[midway,anchor=south, inner sep=1]{\indexsize{$\alpha_8$}};
    \draw[line width=\lw] (V1) -- (R)node[midway,anchor=north west, inner sep=0.1]{\indexsize{$\alpha_2$}} -- (V2)node[midway,anchor=south west, inner sep=0.1]{\indexsize{$\alpha_5$}};
    \draw[line width=\lw] (P2) -- (V1)node[midway,anchor=east, inner sep=0.1]{\indexsize{$\alpha_2$}} -- (P3)node[midway,anchor=south, inner sep=1]{\indexsize{$\alpha_4$}};
    \draw[line width=\lw] (P4) -- (V2)node[midway,anchor=south, inner sep=1]{\indexsize{$\alpha_6$}} -- (P5)node[midway,anchor=east, inner sep=0.1]{\indexsize{$\alpha_7$}};
    \foreach \n/\l in {P1/8,P2/3,P3/4,P4/6,P5/7,P6/10,P7/11}{
        \draw[line width=\lw] (\n) -- ++(0.5*\ttnouterphysdist,0.87*\ttnouterphysdist)node[anchor=-70]{\indexsize{$\sigma_{\l}$}};
    }
    \foreach \n/\l in {R/1,V1/2,V2/5,V3/9}{
        \draw[line width=\lw, dashed] (\n) -- ++(0.5*\ttnouterphysdist,0.87*\ttnouterphysdist)node[anchor=-70]{\indexsize{$\sigma_{\l}$}};
    }
\end{tikzpicture}
        \caption{T3NS}
        \label{fig:t3ns}
    \end{subfigure}
    \begin{subfigure}{0.49\textwidth}
        \begin{tikzpicture}
    \useasboundingbox (-3.5*\ttnvirtdist,-2.5*\ttnvirtdist) rectangle (2.5*\ttnvirtdist,2.5*\ttnvirtdist);

    \node[operatornode] (R) at (0,0){\indexsize{$1$}};
    \node[operatornode] (V1) at (\invsqrttwo*\ttnvirtdist,\invsqrttwo*\ttnvirtdist) {\indexsize{$2$}};
    \node[operatornode] (P2) at ($(V1) + (0,\ttnvirtdist)$) {\indexsize{$3$}};
    \node[operatornode] (P3) at ($(V1) + (\ttnvirtdist,0)$) {\indexsize{$4$}};
    \node[operatornode] (V2) at (\invsqrttwo*\ttnvirtdist,-1*\invsqrttwo*\ttnvirtdist) {\indexsize{$5$}};
    \node[operatornode] (P4) at ($(V2) + (\ttnvirtdist,0)$) {\indexsize{$6$}};
    \node[operatornode] (P5) at ($(V2) + (0,-1*\ttnvirtdist)$) {\indexsize{$7$}};
    \node[operatornode] (P1) at ($(R) + (-1*\ttnvirtdist,0)$) {\indexsize{$8$}};
    \node[operatornode] (V3) at ($(P1) + (-1*\ttnvirtdist,0)$) {\indexsize{$9$}};
    \node[operatornode] (P6) at ($(V3) + (-1*\invsqrttwo*\ttnvirtdist,\invsqrttwo*\ttnvirtdist)$) {\indexsize{$10$}};
    \node[operatornode] (P7) at ($(V3) + (-1*\invsqrttwo*\ttnvirtdist,-1*\invsqrttwo*\ttnvirtdist)$) {\indexsize{$11$}};
    \draw[line width=\lw] (P6) -- (V3) --(P7);
    \draw[line width=\lw] (V3) -- (P1) -- (R);
    \draw[line width=\lw] (V1) -- (R) -- (V2);
    \draw[line width=\lw] (P2) -- (V1) -- (P3);
    \draw[line width=\lw] (P4) -- (V2) -- (P5);
    \foreach \n/\l in {P1/7,P2/2,P3/3,P4/5,P5/6,P6/9,P7/10}{
        \draw[line width=\lw] (\n) -- ++(0.5*\ttnouterphysdist,0.87*\ttnouterphysdist);
        \draw[line width=\lw] (\n) -- ++(-0.5*\ttnouterphysdist,-0.87*\ttnouterphysdist);
    }
    \foreach \n/\l in {R/0,V1/1,V2/4,V3/8}{
        \draw[line width=\lw, dashed] (\n) -- ++(0.5*\ttnouterphysdist,0.87*\ttnouterphysdist);
        \draw[line width=\lw, dashed] (\n) -- ++(-0.5*\ttnouterphysdist,-0.87*\ttnouterphysdist);
    }
\end{tikzpicture}
        \caption{T3NO}
        \label{fig:t3no}
    \end{subfigure}
    \caption{Diagrammatic representation of the different tree tensor networks considered in this work. The dashed lines denote a trivial leg, i.e., one of dimension $1$. The upper two tensor networks are also known as matrix product states and matrix product operators (also called tensor-trains), while the lower two are a variant of trees known as T3NS. The two TTNs on the left represent quantum states, i.e., they are TTNS, while the two on the right represent quantum operators, i.e., TTNOs. The labels on the latter's legs were omitted for visual clarity.}
\end{figure*}
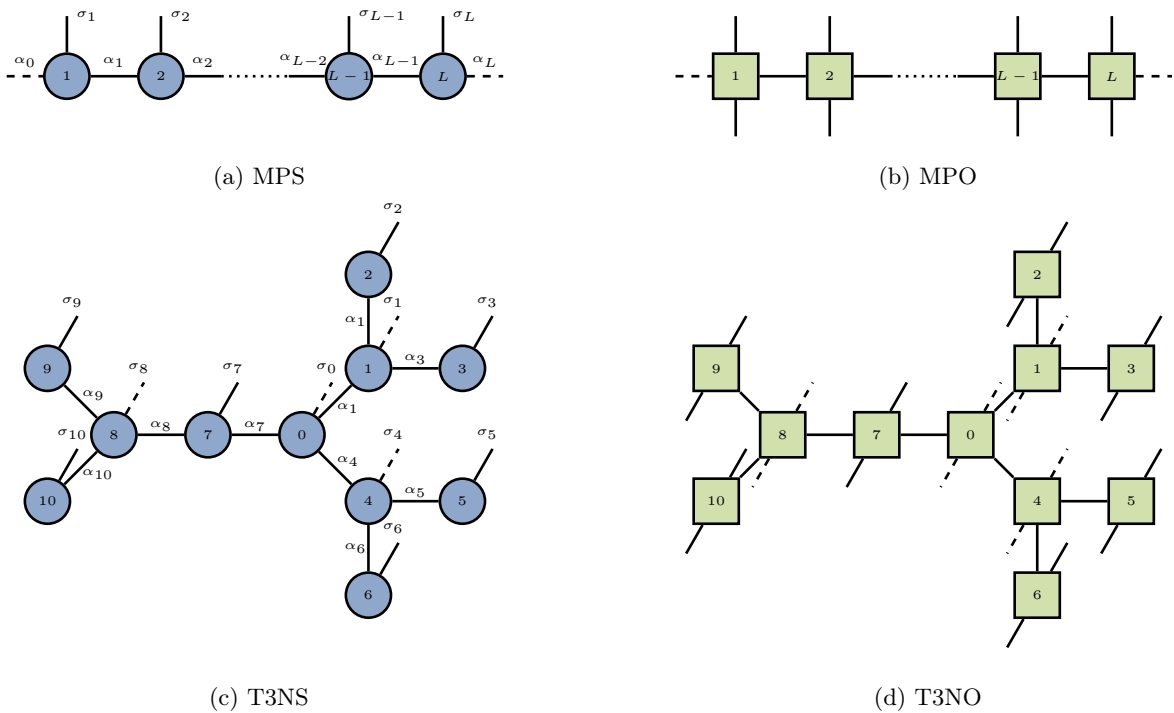

Rather than representing a quantum state $\ket{\psi} \in \C^{d_1 \times \cdots \times d_L}$ of $L$ sites, where site $i$ has local dimension $d_i$, as a state vector, we can describe it as a tree tensor network state (TTNS)~\cite{Shi2006}. Thus, we write the state as
\begin{equation}\label{eq:ttns}
    \ket{\psi} = \sum_{\vec{\sigma},\vec{\alpha}} T^{[1]}_{\sigma_1 \vec{\alpha}_1} \cdots T^{[L]}_{\sigma_L \vec{\alpha}_L} \ket{\sigma_1 \cdots \sigma_L},
\end{equation}
where $\{ \ket{\sigma_i} \}_{\sigma_i=1}^{d_i}$ is the local basis of site $i$, where $d_i$ can be $1$. We also refer to $\sigma_i$ as the physical leg of the tensor $T^{[i]}$. The other legs are virtual legs, where $\alpha_{i,j}=\alpha_{j,i}$ connects the two tensors $T^{[i]}$ and $T^{[j]}$, and the vector $\vec{\alpha}_i$ is a vector over all virtual legs of tensor $T^{i]}$. This connectivity has an underlying tree structure, hence TTNS. For ease of notation, we will root the underlying tree at some site $r$. This creates a parent $p_i$ and a set of children $C_i$ for every site $i$. Since the parent of each site is unique, we can identify $\alpha_{i,p_i}=\alpha_i$ to shorten notation. An additional notation used for the virtual legs is $\vec{\alpha}_{i\setminus k}$, which denotes all the legs to the neighbours of site $i$ with the exclusion of $\alpha_{i,k}$.

In our text, the choice of $r$ is arbitrary and is used merely to facilitate the description of the following algorithms. It is trivial to extend all of them to non-rooted trees. Finally, note that we do not make a clear distinction between a given leg of a tensor and its index values in a contraction sum to simplify the required notation. We will now examine a few special TTNS used throughout this work.

\subsubsection{Special Tree Structures}
The matrix product state (MPS) \cite{Schollwock2011}, also known as a tensor train \cite{Oseledets2011}, is the most commonly used tensor network to represent a quantum state. It is a chain of tensors with two virtual legs and one physical leg each, where the one virtual leg of the first and last tensor is trivial. This structure can be viewed as a TTNS, where every site has at most one child. We will use a slightly different notation compared to \autoref{eq:ttns} and represent a quantum state $\ket{\psi} \in \C^{d_1 \times \cdots \times d_L}$ of $L$ sites by
\begin{equation}
    \ket{\psi} = \sum_{\vec{\sigma}, \vec{\alpha}} T^{[1]}_{\sigma_1 \alpha_0 \alpha_1} \cdots T^{[L]}_{\sigma_L \alpha_{L-1} \alpha_L} \ket{\sigma_1 \cdots \sigma_L}
\end{equation}
where the outermost legs $\alpha_0$ and $\alpha_L$ are of dimension $1$. Note that this notation is consistent with the TTNS notation if we choose an imaginary site $0$ as the root of the underlying tree, potentially with a scalar tensor of value $1$. Since the MPS has been so widely used in research over the last two decades, one cannot hope to state all its use cases here, which span condensed matter physics, quantum information, open quantum systems, and quantum chemistry. A graphical depiction of the MPS structure is given in ~\autoref{fig:mps}. Note that in this work, we will refer to the underlying structure as the tensor train structure, but when used to represent a quantum state, we will refer to the state as an MPS.

The second important tree structure is the T3NS. A T3NS is a tree tensor network state for which no node has more than three non-trivial legs \cite{Gunst2018}. Notably, this means that only a site with at most two non-trivial virtual legs can also have a non-trivial physical leg. This limitation of legs tends to make algorithm scaling acceptable while still allowing full flexibility in creating topologically adapted tree structures. However, especially in cases where physical dimensions are small, it can sometimes be advisable to allow multiple open legs per tensor, while still restricting the number of virtual legs, which tend to have a much higher dimension~\cite{Tagliacozzo2009, Cheng2019, Seitz2023, Dubey2025, Krinitsin2025}. Thus, T3NS have mainly been used in quantum chemistry \cite{Gunst2018, Gunst2019, Ke2025, Ke2025B, Sun2025} or in a setting where neighboring tensors do not both have physical legs~\cite{Silvi2010, Magnifico2021, Ceruti2021, Okunishi2023, Ceruti2023, Sulz2024}. An example of such a T3NS is depicted in ~\autoref{fig:t3ns}.

These are the two main tree structures needed for now. We will explain additional structures as needed, though we will mostly refrain from drawing the leg indices in the tensor network diagrams to avoid unnecessary clutter.

\subsubsection{Tree Tensor Network Operators}
Another kind of tensor network, essential for this work, is the tree tensor network operator (TTNO)~\cite{Frowis2010, Milbradt2024}. Analogously to TTNS representing quantum states, TTNOs represent linear operators acting on them. Thus a linear operator $\hat{O} : \C^{\times d_i} \rightarrow \C^{\times d_i}$ acting on a multi-site quantum system can be written as
\begin{align}\label{eq:ttno}
    \hat{O} = \sum_{\vec{\sigma}', \vec{\sigma}, \vec{\beta}} & \mathcal{T}^{[1]}_{\sigma'_1 \sigma_1 \vec{\beta}_1} \cdots \mathcal{T}^{[L]}_{\sigma'_L \sigma_L \vec{\beta}_L} \nonumber \\ 
    \cdot & \ket{\sigma'_1 \cdots \sigma'_L}\bra{\sigma_1 \cdots \sigma_L}.
\end{align}
This resembles the definition of the TTNS in \autoref{eq:ttns}, apart from the additional physical leg on each tensor. We will also use the same notation for the legs and indices of TTNO tensors, though exchanging $\alpha$ for $\beta$. Both an MPO, the tensor train representation of a TTNO, and a T3NO, the operator equivalent of a T3NS, are depicted in ~\autoref{fig:mpo} and ~\autoref{fig:t3no}, respectively. Some common algorithms that involve TTNO construction are ground-state search \cite{Murg2010, Nakatani2013} and time-evolution algorithms \cite{Bauernfeind2020, Ceruti2021, Ren2022, Ceruti2023, Ceruti2024}. In these algorithms, the TTNO represents the system's Hamiltonian, which can be constructed efficiently for arbitrary tree structures \cite{Milbradt2024, Li2024, Cakir2025}. Most algorithms that require applying MPO to MPS are easily generalizable to more complex tree structures. 

\subsection{Successive Deterministic Compression}\label{sec:CBC}
We will now state the general idea of our new compression algorithm, coined the successive deterministic compression (SDC).
The idea is most easily illustrated by an MPS, which we discuss first. It is also closely related to the original DMRG sweep algorithm~\cite{white1993density}. 

Any MPS can be split around a site $i$ into the form
\begin{align}
|\psi\rangle = \sum_{i\alpha_{i-1}\alpha_i} {T}^{[i]}_{\sigma_i \alpha_{i-1} \alpha_{i}} |\sigma_i\rangle |L_{\alpha_{i-1}}\rangle |R_{\alpha_i}\rangle \label{eq:partition}
\end{align}
where the left states are defined as
\begin{align}
|L_{\alpha_{i}}\rangle & = \sum_{\sigma, \alpha} {T}^{[1]}_{\sigma_1\alpha_p\alpha_1} \cdots T^{[i]}_{\sigma_i \alpha_{i-1}\alpha_i}  |\sigma_1 \cdots \sigma_i \rangle \notag \\
& = \sum_{\sigma_i, \alpha_{i-1}} T^{[i]}_{\sigma_i \alpha_{i-1}\alpha_i} |\sigma_i\rangle |L_{\alpha_{i-1}}\rangle
\end{align}
and similarly for the right states. The size of the left basis at each site $i$ is the  bond dimension $D$ of the MPS, and defining either the set of left states at every partition (or the right states at every partition) is equivalent to defining the MPS. 

Imagine that we have a procedure to construct a compressed left basis $\{ |L_{\chi_{i-1}} \rangle \}$ at each partition with a reduced dimension $\bar{D}$, i.e. in Eq.~(\ref{eq:partition}), $\alpha_{i-1} \to \chi_{i-1}$ where the dimension of $\chi_{i-1}$ is $\bar{D}$. 
Then, these can only be entangled with at most $\bar{D}$ states in the space of $\{ |\sigma_i R_{\alpha_i}\rangle\}$. We can find these states from the QR decomposition  \begin{align}
T^{[i]}_{\sigma_i \chi_{i-1}  \alpha_i} = \sum_{\chi_{i-1}'} R^{[i]}_{\chi_{i-1}\chi_{i-1}'} Q^{[i]}_{\sigma_i \chi_{i-1}'\chi_i} \label{eq:compressedqr}
\end{align}
and the $Q^{[i]}$ tensor now defines a compressed right basis $\{ |R_{\chi_i}\rangle\} $ of dimension $\bar{D}$. Then, repeating this procedure, the set of compressed right states at each partition defines the new compressed MPS.

The remaining question is how to obtain the initial set of compressed left states. Note that, since these are only used as the environment for computing the compressed right states (from which we obtain the compressed MPS), they do not need to be compressed very accurately. Thus, to obtain the compressed left states, we perform sequential SVD compressions on the tensors from $T^{[1]}$ to $T^{[L]}$, similar to the warm-up sweep in early DMRG algorithms.

While the above algorithm has been described for an MPS, it is straightforward to extend it to a TTNS. This is because a TTNS can also always be split into two disconnected subsystems $A$ and $B$ by cutting a bond, similar to an MPS, and one can similarly build the compressed environment tensors around any site using a warmup sweep.

As discussed in the introduction, our main application of TN compression will be in the subroutine of applying a quantum operator to a quantum state. Below, we give a more explicit description of the SDC compression algorithm in the context of $\hat{O}|\psi\rangle$ for both MPS/tensor train and TTNS. 

\subsubsection{Tensor Train Structure}\label{sec:cbc_tt}

\begin{figure*}
    \centering
    % \phantomsubcaption % creates a subfigure counter context

  \begin{overpic}[width=0.4\textwidth]{\figpath 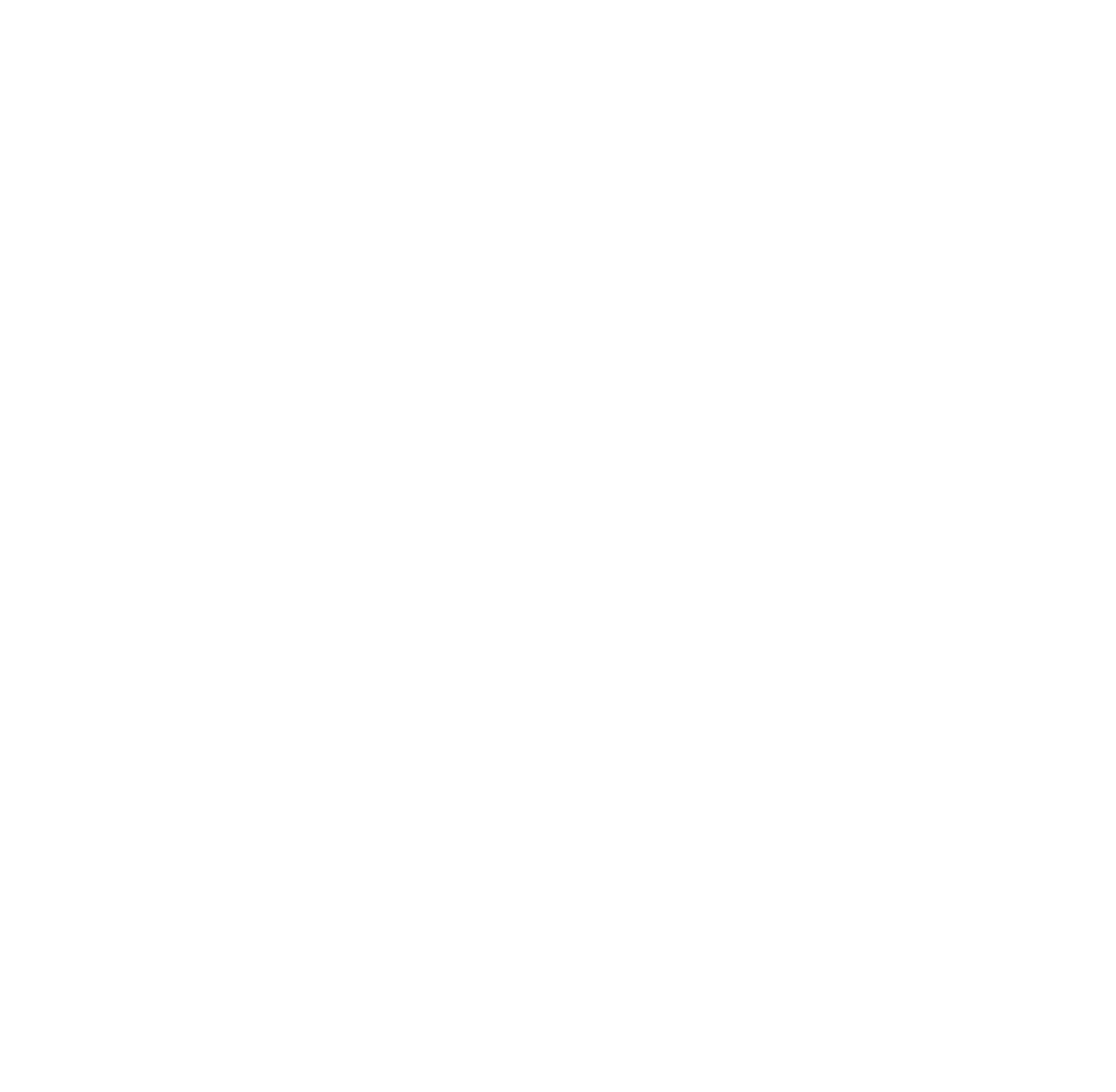}
    \put(-75,0){    \begin{tikzpicture}

        \begin{scope}
            \node[statenode] (S) at (0,0) {};
            \node[operatornode] (O) at (0,\mpsinnerphysdist) {};
            \def\submpsxpos{-1*\mpsvirtdist}
            \draw[line width=\lw] ($(O)+(\submpsxpos, 0)$) -- (O)node[pos=0.65,above]{$\beta_{i-1}$} -- ++(\mpsoutervirtdist,0)node[midway,above]{$\beta_i$};
            \draw[line width=\lw] ($(S)+(\submpsxpos, 0)$) -- (S)node[pos=0.7,above]{$\alpha_{i-1}$} -- ++(\mpsoutervirtdist,0)node[midway,above]{$\alpha_i$};
            \node[submps] (L) at (\submpsxpos,0.5*\mpsinnerphysdist) {};
            \draw[line width=\lw] (S) -- (O)node[midway,right]{$\sigma_i$} -- ++(0,\mpsouterphysdist)node[pos=0.9,right]{$\sigma'_i$};
            \draw[line width=\lw] (L.north) -- ($(O) + (\submpsxpos,\mpsouterphysdist)$)node[pos=0.9,right]{$\chi_{i-1}$};
        \end{scope}

        \begin{scope}[shift={(2.75*\mpsvirtdist,0)}]
            \node[submps] (L) at (0,0.5*\mpsinnerphysdist) {};
            \draw[line width=\lw] ($(L.110) + (0.2mm,-0.2mm)$) -- ++(0,\mpsouterphysdist)node[pos=0.9,left]{$\chi_{i-1}$};
            \draw[line width=\lw] ($(L.70) + (-0.2mm,-0.2mm)$) -- ++(0,\mpsouterphysdist)node[pos=0.9,right]{$\sigma'_i$};
            \draw[line width=\lw] ($(L.60)$) -- ++(\mpsoutervirtdist,0)node[pos=0.6,above]{$\beta_i$};
            \draw[line width=\lw] ($(L.300)$) -- ++(\mpsoutervirtdist,0)node[pos=0.6,above]{$\alpha_i$};
            \draw[tensorcut] ($(L.110) + (-3mm,4mm)$) to[out=-90,in=150] ($(L.110) + (0.2mm,-0.2mm)$) to[out=-60,in=240] ($(L.70) + (-0.2mm,-0.2mm)$) to[out=30,in=-90] ($(L.70) + (3mm,4mm)$);
        \end{scope}

        % Arrow between step 1 and 2
        \def\arrowdist{5mm}
        \draw[intertnarrow] ($(S) + (\mpsoutervirtdist+\arrowdist,0.5*\mpsinnerphysdist)$) -- ($(L) + (-1*\arrowdist-5mm,0)$) node[midway,above]{$\mathcal{C}$};

        \begin{scope}[shift={(5.25*\mpsvirtdist,0)}]
            \node[submps] (L1) at (0,0.5*\mpsinnerphysdist) {};
            \node[isonode,rotate=-90,fill=myellow!40] (iso) at ($(L1.north) + (0,\mpsouterphysdist)$) {};
            \draw[line width=\lw] ($(L1.60)$) -- ++(\mpsoutervirtdist,0)node[midway,above]{$\beta_i$};
            \draw[line width=\lw] ($(L1.300)$) -- ++(\mpsoutervirtdist,0)node[midway,above]{$\alpha_i$};
            \draw[line width=\lw] (L1.north) -- (iso)node[midway,right]{$\chi_i$};
            \draw[line width=\lw] ($(iso.150) + (0.2mm,-0.2mm)$) -- ++(0,\mpsouterphysdist-3mm)node[pos=0.6,left]{$\chi_{i-1}$};
            \draw[line width=\lw] ($(iso.30) + (-0.2mm,-0.2mm)$) -- ++(0,\mpsouterphysdist-3mm)node[pos=0.6,right]{$\sigma'_i$};
        \end{scope}

        % Arrow between step 2 and 3
        \draw[intertnarrow] ($(L) + (\mpsoutervirtdist+\arrowdist,0)$) -- ($(L1) + (-1*\arrowdist-3mm,0)$) node[midway,above]{SVD};

        \begin{scope}[shift={(0,-4*\mpsouterphysdist)}]
            \begin{scope}
                \node[statenode] (S) at (0,0) {};
                \node[operatornode] (O) at (0,\mpsinnerphysdist) {};
                \def\submpsxpos{-1*\mpsvirtdist}
                \def\rightsubmpsxpos{\mpsvirtdist}
                \draw[line width=\lw] ($(O)+(\submpsxpos, 0)$) -- (O)node[pos=0.65,above]{$\beta_{i-1}$} -- ($(O)+(\rightsubmpsxpos, 0)$)node[pos=0.4,above]{$\beta_i$};
                \draw[line width=\lw] ($(S)+(\submpsxpos, 0)$) -- (S)node[pos=0.7,above]{$\alpha_{i-1}$} -- ($(S)+(\rightsubmpsxpos, 0)$)node[pos=0.4,above]{$\alpha_i$};
                \node[submps] (L) at (\submpsxpos,0.5*\mpsinnerphysdist) {};
                \node[submpsright] (R) at (\rightsubmpsxpos,0.5*\mpsinnerphysdist) {};
                \draw[line width=\lw] (S) -- (O)node[midway,right]{$\sigma_i$} -- ++(0,\mpsouterphysdist)node[pos=0.9,right]{$\sigma'_i$};
                \draw[line width=\lw] (L.north) -- ($(O) + (\submpsxpos,\mpsouterphysdist)$)node[pos=0.9,right]{$\chi_{i-1}$};
                \draw[line width=\lw] (R.north) -- ($(O) + (\rightsubmpsxpos,\mpsouterphysdist)$)node[pos=0.9,right]{$\chi_i$};
            \end{scope}

            \begin{scope}[shift={(3*\mpsvirtdist,0.5*\mpsinnerphysdist)}]
                \node[statenode,fill=morange!40] (S) at (0,0) {};
                \draw[line width=\lw] (S) -- ++(-1*\mpsoutervirtdist,0)node[midway,above]{$\chi_{i-1}$};
                \draw[line width=\lw] (S) -- ++(\mpsoutervirtdist,0)node[midway,above]{$\chi_i$};
                \draw[line width=\lw] (S) -- ++(0,\mpsoutervirtdist)node[pos=0.7,right]{$\sigma'_i$};
                \draw[tensorcut] ($(S) + (-0.5*\mpsinnerphysdist,0.6*\mpsouterphysdist)$) to[out=0,in=90] (S.center) to[out=-90,in=0] ++(-0.5*\mpsinnerphysdist,-0.6*\mpsouterphysdist);
            \end{scope}

            % Arrow between step 1 and 2
            \draw[intertnarrow] ($(R) + (1mm+\arrowdist,0)$) -- ($(S) + (-1*\arrowdist-1*\mpsoutervirtdist+1mm,0)$) node[midway,above]{$\mathcal{C}$};

            \begin{scope}[shift={(5.25*\mpsvirtdist,0.5*\mpsinnerphysdist)}]
                \node[matrix, fill=myellow!40] (R) at (0,0) {};
                \node[isonode, rotate=180, fill=morange!40] (Q) at (\mpsvirtdist,0) {};
                \draw[line width=\lw] ($(R) + (-\mpsoutervirtdist,0)$) -- (R)node[midway,above]{$\chi_{i-1}$} -- (Q)node[pos=0.6,above]{$\chi'_{i-1}$} -- + (\mpsoutervirtdist-2mm,0)node[midway,above]{$\chi_i$};
                \draw[line width=\lw] (Q) -- ++(0,\mpsouterphysdist)node[pos=0.7,right]{$\sigma'_i$};
            \end{scope}

            % Arrow between step 2 and 3
            \draw[intertnarrow] ($(S) + (\arrowdist+\mpsoutervirtdist-2mm,0)$) -- ($(R) + (-1*\arrowdist-1*\mpsoutervirtdist+3mm,0)$) node[midway,above]{QR};
    
        \end{scope}

        \begin{scope}[shift={(8*\mpsvirtdist,-1*\mpsinnerphysdist)}]
            \node[statenode] (S) at (0,0) {};
            \node[operatornode] (O) at (0,\mpsinnerphysdist) {};
            \def\rightsubmpsxpos{\mpsvirtdist}
            \draw[line width=\lw] ($(O)+(-1*\mpsoutervirtdist, 0)$) -- (O)node[pos=0.3,above]{$\beta_{i-1}$} -- ($(O)+(\rightsubmpsxpos, 0)$)node[midway,above]{$\beta_i$};
            \draw[line width=\lw] ($(S)+(-1*\mpsoutervirtdist, 0)$) -- (S)node[pos=0.3,above]{$\alpha_{i-1}$} -- ($(S)+(\rightsubmpsxpos, 0)$)node[midway,above]{$\alpha_i$};
            \node[submpsright] (R) at (\rightsubmpsxpos,0.5*\mpsinnerphysdist) {};
            \node[isonode, rotate=90, fill=morange!40] (Q) at (0.5*\mpsvirtdist,2*\mpsinnerphysdist) {$*$};
            \draw[line width=\lw] (O) -- (S)node[midway,right]{$\sigma_i$};
            \draw[line width=\lw] (R.north) to[out=90,in=-90] node[pos=0.3,right]{$\chi_i$}(Q.north);
            \draw[line width=\lw] (O.north) to[out=90,in=180] node[pos=0.3,right]{$\sigma_i$}(Q.east);
            \draw[line width=\lw] (Q) -- ++(0,\mpsouterphysdist)node[pos=0.8,right]{$\chi_{i-1}$};
            \def\equaldist{3mm}
            \def\secondrightpos{\rightsubmpsxpos+\equaldist+\mpsoutervirtdist-3mm}
            % Arrow between step 1 and 2
            \draw[intertnarrow] ($(R) + (\mpsoutervirtdist-3mm,0)$) -- ($(R) + (\arrowdist+\mpsoutervirtdist-1mm,0)$) node[midway,above]{$\mathcal{C}$};
            \node[submpsright] (R2) at ($(R.east) + (\secondrightpos,0)$) {};
            \draw[line width=\lw] ($(R2.120)$) -- ++(-1*\mpsoutervirtdist,0)node[midway,above]{$\beta_{i-1}$};
            \draw[line width=\lw] ($(R2.240)$) -- ++(-1*\mpsoutervirtdist,0)node[midway,above]{$\alpha_{i-1}$};
            \draw[line width=\lw] (R2.north) -- ($(O) + ($(R.east) + (\secondrightpos,0)$)$)node[midway,right]{$\chi_{i-1}$};
        \end{scope}

        % Rectangle upper left
        \draw[line width=\lw] (-1.5*\mpsvirtdist,-0.5*\mpsouterphysdist) rectangle (7*\mpsvirtdist,4*\mpsouterphysdist);
        % \node[anchor=north west] at (-1.5*\mpsvirtdist,4*\mpsouterphysdist) {a)};
        % Rectangle lower left
        \draw[line width=\lw] (-1.5*\mpsvirtdist,-0.5*\mpsouterphysdist) rectangle (7*\mpsvirtdist,-4.5*\mpsouterphysdist);
        % \node[anchor=north west] at (-1.5*\mpsvirtdist,-0.5*\mpsouterphysdist) {b)};
        % Rectangle right
        \draw[line width=\lw] (7*\mpsvirtdist,4*\mpsouterphysdist) rectangle (11.7*\mpsvirtdist,-4.5*\mpsouterphysdist);
        % \node[anchor=north west] at (7*\mpsvirtdist,4*\mpsouterphysdist) {c)};

    \end{tikzpicture}}

    % (a)
    \put(-73,97){(a)\phantomsubcaption\label{fig:mps_cbc_a}}

    % (b)
    \put(-73,43){(b)\phantomsubcaption\label{fig:mps_cbc_b}}

    % (c)
    \put(86,97){(c)\phantomsubcaption\label{fig:mps_cbc_c}}
  \end{overpic}
    \caption{The different steps of the SDC for a tensor train structure. a) The contraction \autoref{eq:tt_cbc_first_contraction} and subsequent truncation, b) the contraction \autoref{eq:right_sweep_init_contr} and subsequent QR-decomposition \autoref{eq:tt_cbc_qr_decomp} to obtain the new local tensor, c) the final contraction \autoref{eq:tt_cbc_final_contraction} to find the new projected subsystem tensor used for the next site. Note that the yellow tensors are not used in later steps and can be deleted.}
    \label{fig:mps_cholesky_compression}
\end{figure*}
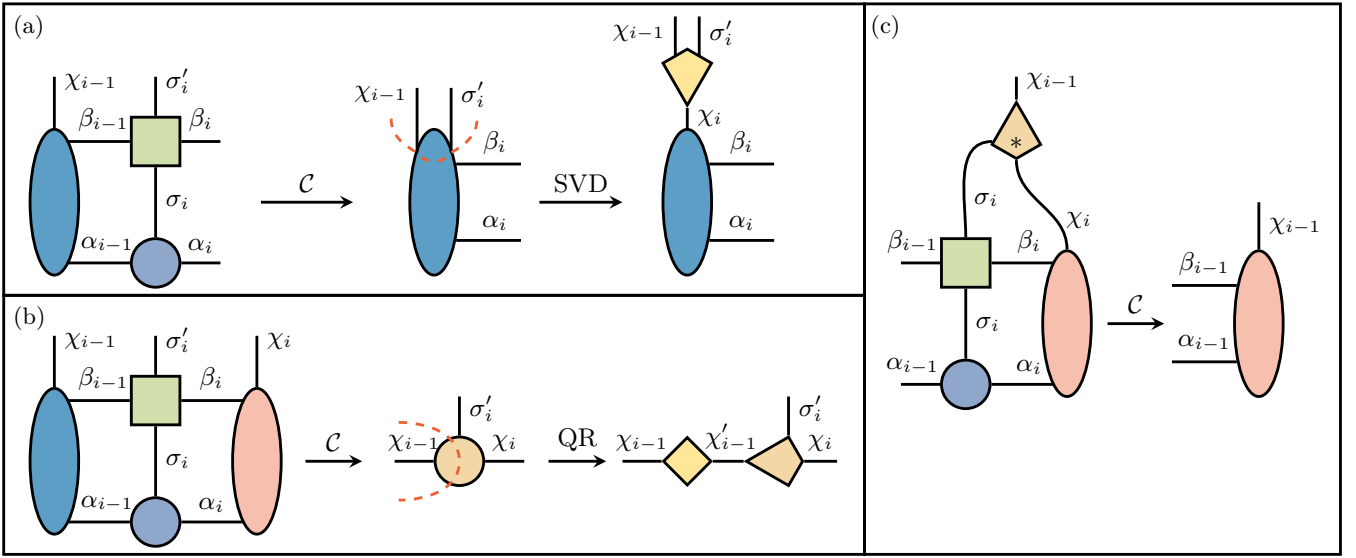

For the chain-like tensor train structure, start from the leftmost site and obtain a left environment for every site. To do so, we start performing the contraction
\begin{equation}\label{eq:tt_cbc_first_contraction}
    \mathcal{L}'^{[i]}_{\chi_{i-1}\sigma'\alpha_{i}\beta_{i}} = \sum_{\substack{\sigma_i\\ \alpha_{i-1}\beta_{i-1}}} \mathcal{L}^{[i-1]}_{\chi_{i-1}\alpha_{i-1}\beta_{i-1}}\mathcal{T}_{\sigma_i'\sigma_i\beta_{i-1}\beta_{i}}^{[i]} T_{\sigma_i\alpha_i\alpha_{i-1}}^{[i]}
\end{equation}
for site $i$ with $\mathcal{L}^{[0]} = 1$ as a an initial condition. The leg denoted by $\chi$ corresponds to the current set of (compressed) left states for the sites up to $i-1$. To obtain the compressed left states $\chi_i$ for the current site $i$, the legs $\chi_{i-1}$ and $\sigma'$ of $\mathcal{L}'^{[i]}$ are combined into one leg $(\chi_{i-1},\sigma')$. The newly obtained leg $(\chi_{i-1},\sigma')$ is then compressed down to the desired new bond dimension $\Bar{D}$, via singular value decomposition. This yields the new leg $\chi_i$ and the tensor $\mathcal{L}^{[i]}$, which is saved for later. Any isometric tensor on the other end of leg $\chi_i$, i.e., a tensor that now has the large leg $(\chi_{i-1},\sigma')$, can be dropped. A diagrammatic version of this procedure is shown in ~\autoref{fig:mps_cbc_a}. We do this for all sites except the rightmost one, $i=L$, for which no action is needed in this sweep.

Once we reach the rightmost site, we perform a backwards sweep to construct the tensors that lead to the compressed MPS. We follow these steps for each site. First, we perform the contraction
\begin{align}\label{eq:right_sweep_init_contr}
    \mathcal{S}^{[i]}_{\sigma'_i \chi_{i-1} \chi_i} = \sum_{\substack{\sigma\\\alpha_{i-1}\beta{i-1}\\\alpha_i\beta_i}} & \mathcal{L}^{[i-1]}_{\chi_{i-1}\alpha_{i-1}\beta_{i-1}}\mathcal{T}_{\sigma_i'\sigma_i\beta_i\beta_{i-1}}^{[i]} \nonumber \\
    & \cdot T_{\sigma_i\alpha_i\alpha_{i-1}}^{[i]} \mathcal{R}^{[i]}_{\chi_i\alpha_i\beta_i}
\end{align}
with the initial condition $\mathcal{R}^{[L]}=1$. If we are at the leftmost site $i=1$, $\mathcal{S}^{[1]}$ will be the new MPS tensor $T'^{[1]}$. Otherwise, we continue and perform a QR-decomposition on leg $\chi_{i-1}$ of $\mathcal{S}^{[i]}$ to obtain
\begin{equation}\label{eq:tt_cbc_qr_decomp}
    \mathcal{S}^{[i]}_{\sigma'_i \chi_{i-1} \chi_i} = \sum_{\chi'_{i-1}}R_{\chi_{i-1}\chi'_{i-1}} Q^{[i]}_{\sigma'_i \chi'_{i-1} \chi_i}
\end{equation}
of which we assign the isometric tensor $Q^{[i]}$ as the new MPS tensor $T'^{[i]}$ of this site. This part of the procedure is shown in ~\autoref{fig:mps_cbc_b}. We then obtain the new tensor $\mathcal{R}^{[i-1]}$ by performing the contraction
\begin{align}\label{eq:tt_cbc_final_contraction}
    \mathcal{R}^{[i-1]}_{\chi'_{i-1}\alpha_{i-1}\beta_{i-1}} =  \sum_{\substack{\sigma'_i\sigma_i\chi_i\\ \alpha_i\beta_i}} & Q^{[i]*}_{\sigma'_i \chi'_{i-1} \chi_i} T^{[i]}_{\sigma_i\alpha_{i-1}\alpha_i} \nonumber \\
     \cdot & \mathcal{T}^{[i]}_{\sigma'_i\sigma_i\beta_{i-1}\beta_i} \mathcal{R}^{[i]}_{\chi_i\alpha_i\beta_i}.
\end{align}
This contraction is shown graphically in ~\autoref{fig:mps_cbc_c}. Then we continue with the next site. To reiterate, once the leftmost site is reached, we merely perform the contraction in \autoref{eq:right_sweep_init_contr}, and we will have obtained our new MPS tensors as
\begin{equation}
    T'^{[i]} = \begin{cases*}
        \mathcal{S}^{[i]} & if $i=1$ \\
        Q^{[i]} & else.
    \end{cases*}
\end{equation}
A detailed run-through of this process for an MPS of length $4$ is shown in App.~\autoref{app:mps_illustration}.

\subsubsection{Tree Structures}\label{sec:cbc_tree}

\begin{figure*}
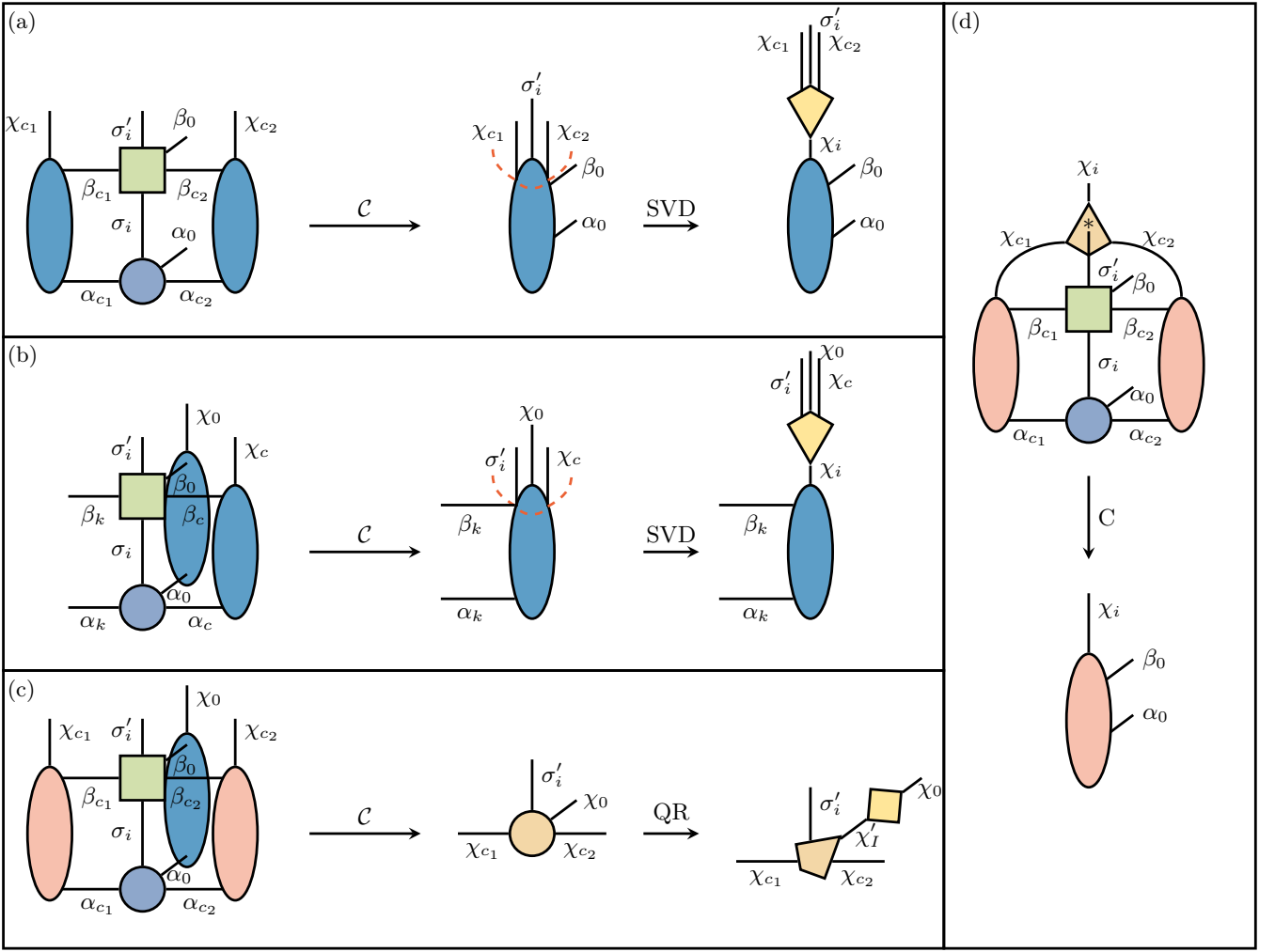

    \centering

    \begin{overpic}[width=0.8\textwidth]{\figpath white.png}
    \put(-13,0){\includestandalone[width=\textwidth]{\tikzpath ttns_cbc}}

    % (a)
    \put(-12,91){(a)\phantomsubcaption\label{fig:ttns_cbc_a}}

    % (b)
    \put(-12,58){(b)\phantomsubcaption\label{fig:ttns_cbc_b}}

    % (c)
    \put(-12,25){(c)\phantomsubcaption\label{fig:ttns_cbc_c}}

    % (d)
    \put(81,91){(d)\phantomsubcaption\label{fig:ttns_cbc_d}}
  \end{overpic}
    \caption{The different steps of the SDC for a T3NS structure.  a) The contraction and truncation from leaves to root \autoref{eq:cbc_tree_first_contraction}, b) the contraction and truncation from the root to the leaves \autoref{eq:cbc_tree_second_contraction}, c) The contraction \autoref{eq:cbc_tree_third_contraction} and subsequent QR-decomposition \autoref{eq:cbc_tree_qr} to obtain the new local tensor, d) the final contraction \autoref{eq:cbc_final_contraction} to obtain the new projected subsystem tensor for the parent site. Note that the yellow tensors are not used in later steps and can be deleted.}
    \label{fig:ttns_cholesky_compression}
\end{figure*}

We now extend the SDC algorithm to tensor networks with a tree structure. The graphical depiction of a single step in T3NS is shown in ~\autoref{fig:ttns_cholesky_compression}. Overall, the algorithm works analogously to the tensor train case, though it splits into three different main steps. First, the subtree tensors are constructed from the leaves to the root, then from the root to the leaves. These are the analogues of the compressed left (environment) states in the tensor train/MPS example. Finally, we extract the compressed TTNS representation by performing QR decompositions from the leaves to the root.

Thus, first, we move up from the leaves to the root to construct the subtree tensors. For each site $i$, we perform the contraction
\begin{equation}\label{eq:cbc_tree_first_contraction}
    \mathcal{L}'^{[i]}_{\sigma'_i \vec{\chi}_{i\setminus 0} \alpha_p \beta_p} = \sum_{\substack{\sigma_i \\ \vec{\alpha}_{i\setminus 0} \vec{\beta}_{i\setminus 0}}} T^{[i]}_{\sigma_i\vec{\alpha}_i} \mathcal{T}^{[i]}_{\sigma'_i\sigma_i\vec{\beta}_i} \prod_{c\in C_i} \mathcal{L}^{[c]}_{\chi_c \alpha_c \beta_c}.
\end{equation}
We then join the legs $(\sigma'_i,\vec{\chi}_i)$ into one and compress it as we did in the previous section. This yields the new leg $\chi_i$ with desired bond dimension $D$ and the tensor $\mathcal{L}^{[i]}$. These steps are shown in ~\autoref{fig:ttns_cbc_a}. Once we reach the root of the tree, we perform the following steps for every site $i$ except the leaves, as we move down the tree for the second step. For each $k \in C$ we contract
\begin{align}\label{eq:cbc_tree_second_contraction}
    \mathcal{L}'^{[(i,k)]}_{\sigma'_i\vec{\chi}_{i\setminus k} \alpha_k \beta_k} = \sum_{\substack{\sigma_i\\ \vec{\alpha}_{i\setminus k}\vec{\beta}_{i \setminus k}}} & T^{[i]}_{\sigma_i\vec{\alpha}_i} \mathcal{T}^{[i]}_{\sigma'_i\sigma_i\vec{\beta}_i} \nonumber \\
    & \cdot \mathcal{L}^{[(p_i,i)]}_{\chi_p\alpha_p\beta_p} \prod_{c\in C_i \setminus k} \mathcal{L}^{[c]}_{\chi_c \alpha_c \beta_c}.
\end{align}
We combine the legs $(\sigma'_i,\vec{\chi}_{i \setminus k})$ into one and compress it, yielding the leg $\chi_i$ and the tensor $\mathcal{L}^{[(i,k)]}$. The three steps are shown in ~\autoref{fig:ttns_cbc_b}. These first two parts are almost the same as the steps for the tensor train structure, apart from the choice of which legs are contracted.

Now we do another sweep, starting at the leaf of the tree and moving towards the root. For every site $i$, we contract
\begin{align}\label{eq:cbc_tree_third_contraction}
    \mathcal{S}_{\sigma'_i\vec{\chi}_i} = \sum_{\substack{\sigma_i\vec{\alpha}_i \vec{\beta}_i}} & T^{[i]}_{\sigma_i\vec{\alpha}_i} \mathcal{T}^{[i]}_{\sigma'_i\sigma_i\vec{\beta}_i} \nonumber \\
    \cdot & \mathcal{L}^{[(p_i,i)]}_{\chi_p\alpha_p\beta_p} \prod_{c\in C_i} \mathcal{R}^{[c]}_{\chi_c \alpha_c \beta_c},
\end{align}
as shown in the first step of ~\autoref{fig:ttns_cbc_c}. If $i$ is the root site, we are done and consider $T'^{[i]} = \mathcal{S}^{[i]}$ as the new TTNS tensor. Otherwise we continue by performing a QR decomposition on leg $\chi_i$ of $\mathcal{S}^{[i]}$ to obtain
\begin{equation}\label{eq:cbc_tree_qr}
    \mathcal{S}^{[i]}_{\sigma'_i \vec{\chi}_i} = \sum_{\chi'_i} Q^{[i]}_{\sigma_i'\chi_i'\vec{\chi}_i} R_{\chi'_i\chi_i}.
\end{equation}
The resulting isometric tensor $Q^{[i]}$ is then assigned to be the new tensor $T'^{[i]}$ in the TTNS. The decomposition is shown as the second step in ~\autoref{fig:ttns_cbc_c}. As a final step we obtain the new tensor $\mathcal{R}^{[i]}$ via the contraction
\begin{align}\label{eq:cbc_final_contraction}
    \mathcal{R}^{[i]}_{\chi'_0\alpha_p\beta_p} = \sum_{\substack{\sigma'_i\sigma_i\\ \vec{\alpha}_{i \setminus 0}\vec{\beta}_{i \setminus 0}\vec{\chi}_i}} & T^{[i]}_{\sigma_i\vec{\alpha}_i} \mathcal{T}^{[i]}_{\sigma'_i\sigma_i\vec{\beta}_i} \nonumber \\
    \cdot & \, Q^{[i]}_{\sigma_i'\chi_i'\vec{\chi}_i} \prod_{c\in C_i} \mathcal{R}^{[c]}_{\chi_c\alpha_c\beta_c}.
\end{align}
A graphical depiction of this step is given in ~\autoref{fig:ttns_cbc_d}. Then we continue with the parent of this site. So eventually we end up with the new tensors for the TTNS
\begin{equation}
    T'^{[i]} = \begin{cases*}
        \mathcal{S}^{[i]} & if $i$ is the root \\
        Q^{[i]} & else.
    \end{cases*}
\end{equation}

\subsubsection{Other Compression Methods}
We will now provide an overview of other methods commonly used to apply an MPO to an MPS and to compress the resulting bond dimension. For some of these methods, there is no explicit description of a general TTN version in the literature. However, while not trivial to implement, the approach to generalize from an MPS to a TTN is conceptually straightforward, as illustrated in the previous section. We therefore interchangeably use the language of MPS and TTN in the descriptions below.

The simplest way to perform the application and compression is to do them separately. Here, the TTNO is fully contracted with the TTNS, and the two virtual legs (from the TTNO and TTNS) between the two sites are combined into one larger leg. Then the compression is performed as a second step. This compression takes the form of a sweep through the entire network, performing truncated SVDs for each virtual leg encountered during the sweep to compress it down to the desired dimension. The sweep can be performed either from leaves to root, requiring the movement of the orthogonality centre \cite{Oseledets2011, Schollwock2011, Ma2024}, or recursively from the root to the leaves with the root as the orthogonality centre \cite{Ceruti2023}. The compression can also benefit from using randomised SVD if the difference between the current and desired bond dimension is large \cite{AlDaas2023}. While this approach is the most accurate one \cite{Oseledets2011}, it has the highest cost scaling with the TTNO and TTNS bond dimensions, compared to other methods \cite{Paeckel2019, AlDaas2023, Camano2025, Ma2024}. We refer to this approach below as the Direct method.

The other end of the spectrum is the ZipUp method \cite{Stoudenmire2010}. This is very fast but not very accurate. The method sweeps through the network only once and compresses each site locally. There are two reasons for its relative inaccuracy and high speed compared to the Direct method. The first is that the tensor whose leg is being compressed also has other compressed legs yielded by previous compressions, which restrict the information gained from the previously compressed tensors. The second is that the compression is not performed with the TTN being in full canonical form with respect to the resulting TTNS, as the TTNO and TTNS are placed in canonical form independent from each other~\cite{Paeckel2019}. For a graphical depiction of this method for MPS, consider any of \cite{Paeckel2019, Ma2024, tensornetworkorgzipup}.

The density matrix (DM)- based method lies between the Direct and ZipUp methods in terms of accuracy and speed for MPS, although we will see later that the relative complexity differs for general TTN \cite{DMCNote}. This algorithm utilises the relation between the bond dimension in a TTNS and the entanglement of the two subsystems connected by the bonds in question. As discussed earlier, every TTNS can be split into two subsystems along any bond. By sweeping through the TTNS, one can iteratively construct the reduced density matrices of every bond. Then the eigendecomposition is used to truncate the desired bond and obtain a good basis for the subtree, which is similar to the SDC algorithm

A recent algorithm is the successive randomised compression (SRC) \cite{Camano2025}. It utilises multiple randomly generated tensors to find the compressed TN representation, employing the same underlying ideas as random matrix decompositions \cite{Halko2011, Murray2023}. For this algorithm, a single random matrix of size $d \times \Bar{D}$ is drawn per site. They are combined via the Khatri-Rao product and contracted to build environment blocks similar to our SDC algorithm. However, instead of an SVD to obtain compressed environment states, compression is achieved via contraction with random matrices. Performance-wise, the fidelity of the obtained state is comparable to that of the Direct method and DMC algorithms, while beating the Zip-Up algorithm in runtime \cite{Camano2025}. One can view the SRC as the randomised version of our SDC algorithm.

Finally, there is an entire class of algorithms that variationally fit the result of the TTNO applied to the TTNS to a new TTNS of desired bond dimension. This can be done by first contracting and then performing the fitting \cite{Schollwock2011, Paeckel2019}. A way to perform the application without the explicit contraction can be achieved via DMRG-like sweeps \cite{Verstraete2004, Stoudenmire2010}. However, these variational approaches can take a long time to converge or get stuck in local minima without an educated starting guess, for example, obtained via a Zip-Up run \cite{Paeckel2019, Camano2025}. Therefore, we will exclude these approaches from the following numerical exploration.

A simplified scaling of the operation count is presented in ~\autoref{tab:time_scal}, and a simplified scaling for the required memory in ~\autoref{tab:mem_scal}. Because there are many steps in the algorithms, and each step scales differently with the state, operator, and compressed dimension, the table makes some assumptions about the relative sizes of these bond dimensions. In particular, it assumes $d \leq D_S \leq \bar{D}$, $d \leq D_O \leq \bar{D}$, and $\Bar{D} \leq D_SD_O$.

\begin{table}
    \begin{tabular}{c|c|c}
        Computational cost & MPS & T3NS \\
        \hline
        Direct & $\mathcal{O}(LdD_S^3D_O^3)$ & $\mathcal{O}(Ld^2D_S^4D_O^4)$ \\
        Density Matrix & $\mathcal{O}(Ld^2D_SD_O\Bar{D}^2)$ & $\mathcal{O}(Ld D_S^4 D_O^4)$ \\
        Zip-Up & $\mathcal{O}(Ld^2D_SD_O\Bar{D}^2)$ & $\mathcal{O}(LdD_S^2D_O^2\Bar{D}^2)$ \\
        SRC & $\mathcal{O}(LdD_SD_O\Bar{D}^2)$ & $\mathcal{O}(LdD_S^2D_O^2\Bar{D}^2)$ \\
        SDC & $\mathcal{O}(Ld^2D_SD_O\Bar{D}^2)$ & $\mathcal{O}(Ld^2D_SD_O\Bar{D}^4)$ \\
    \end{tabular}
    \caption{The scaling of the operation count of the different application methods for MPSs and T3NSs under the assumptions $d \leq D_S \leq \Bar{D}$, $d \leq D_O \leq \Bar{D}$, and $\Bar{D} < D_SD_O$ where $d$ is the maximum physical dimension, $D_S$ and $D_O$ the maximum virtual bond dimension of the state and operator, respectively, and $\Bar{D}$ the desired bond dimension after the contraction.}
    \label{tab:time_scal}
\end{table}

\begin{table}
    \begin{tabular}{c|c|c}
        Memory cost & MPS & T3NS \\
        \hline
        Direct & $\mathcal{O}(dD_S^2D_O^2)$ & $\mathcal{O}(dD_S^3D_O^3)$ \\
        Density Matrix & $\mathcal{O}(d^2D_S^2D_O^2)$ & $\mathcal{O}(d^2\Bar{D}^4)$ \\
        Zip-Up & $\mathcal{O}(dD_SD_O\Bar{D})$ & $\mathcal{O}(dD_OD_S\Bar{D}^2)$ \\
        SRC & $\mathcal{O}(dD_SD_O\Bar{D})$ & $\mathcal{O}(dD_SD_O\Bar{D}^2)$ \\
        SDC & $\mathcal{O}(dD_SD_O\Bar{D})$ & $\mathcal{O}(dD_SD_O\Bar{D}^2)$ \\
    \end{tabular}
    \caption{The scaling of memory requirements for the different application methods for MPSs and T3NSs under the same assumptions as \autoref{tab:time_scal}.}
    \label{tab:mem_scal}
\end{table}

\section{Results}\label{sec:numerical_sims}
We will now compare our newly introduced SDC to the other application methods. The numerical simulations were performed using PyTreeNet \cite{Milbradt2024B, PyTreeNet} on an Intel i7-10700 CPU with 32 GB of RAM.

% Note that this figure belongs to the next subsection
%Moved here for layout purposes
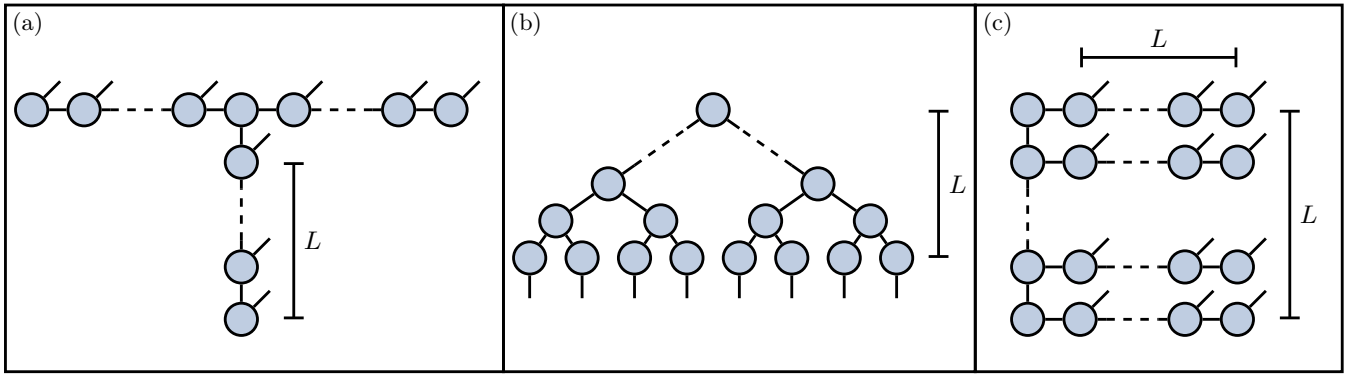
\begin{figure*}
    \centering
    \begin{overpic}[width=0.3\textwidth]{\figpath white.png}
    \put(-115,0){\begin{tikzpicture}
    \def\nodedist{0.65}
    \def\physdist{0.5}

    \node[csimnode] (R) at (0,0) {};
    \foreach \i in {1,3,4}{
        \node[csimnode] (Left\i) at ($(R) + (-1*\i*\nodedist,0)$) {};
        \draw[line width=\lw] (Left\i) -- ++(\invsqrttwo*\physdist,\invsqrttwo*\physdist);
        \node[csimnode] (Right\i) at ($(R) + (\i*\nodedist,0)$) {};
        \draw[line width=\lw] (Right\i) -- ++(\invsqrttwo*\physdist,\invsqrttwo*\physdist);
        \node[csimnode] (Down\i) at ($(R) + (0,-1*\i*\nodedist)$) {};
        \draw[line width=\lw] (Down\i) -- ++(\invsqrttwo*\physdist,\invsqrttwo*\physdist);
    }
    \draw[line width=\lw] (Left4) -- (Left3) -- ++(0.5*\nodedist,0);
    \draw[line width=\lw] (Right4) -- (Right3) -- ++(-0.5*\nodedist,0);
    \draw[line width=\lw] (Down4) -- (Down3) -- ++(0,0.5*\nodedist);
    \draw[line width=\lw, dashed] ($(Left3) + (0.5*\nodedist,0)$) -- ($(Left1) + (-0.5*\nodedist,0)$);
    \draw[line width=\lw, dashed] ($(Right3) + (-0.5*\nodedist,0)$) -- ($(Right1) + (0.5*\nodedist,0)$);
    \draw[line width=\lw, dashed] ($(Down3) + (0,0.5*\nodedist)$) -- ($(Down1) + (0,-0.5*\nodedist)$);
    \draw[line width=\lw] (R) -- (Left1) -- ++(-0.5*\nodedist,0);
    \draw[line width=\lw] (R) -- (Right1) -- ++(0.5*\nodedist,0);
    \draw[line width=\lw] (R) -- (Down1) -- ++(0,-0.5*\nodedist);
    \draw[line width=\lw, |-|] (\nodedist,-1*\nodedist) -- ++(0,-3*\nodedist)node[midway,right]{$L$};

    \begin{scope}[shift={(9*\nodedist,0)}]
        \node[csimnode] (R) at (0,0) {};
        \foreach \i in {0,1,2,3}{
            \node[csimnode] (QL\i) at ($(R) + (-0.5*\nodedist-1*\i*\nodedist,-4*\invsqrttwo*\nodedist)$) {};
            \draw[line width=\lw] (QL\i) -- ++(0,-1*\physdist);
            \node[csimnode] (QR\i) at ($(R) + (0.5*\nodedist+1*\i*\nodedist,-4*\invsqrttwo*\nodedist)$) {};
            \draw[line width=\lw] (QR\i) -- ++(0,-1*\physdist);
        }
        \foreach \i in {1,3}{
            \node[csimnode] (VL1\i) at ($(R) + (-1*\i*\nodedist,-3*\invsqrttwo*\nodedist)$) {};
            \node[csimnode] (VR1\i) at ($(R) + (1*\i*\nodedist,-3*\invsqrttwo*\nodedist)$) {};
        }
        \node[csimnode] (VL2) at ($(R) + (-1*2*\nodedist,-2*\invsqrttwo*\nodedist)$) {};
        \node[csimnode] (VR2) at ($(R) + (2*\nodedist,-2*\invsqrttwo*\nodedist)$) {};
        \foreach \l/\c/\r in {QL3/VL13/QL2, QL1/VL11/QL0, QR0/VR11/QR1, QR2/VR13/QR3, VL13/VL2/VL11, VR11/VR2/VR13}{
            \draw[line width=\lw] (\l) -- (\c) -- (\r);
        }
        \draw[line width=\lw] (VL2) -- ++(0.5*\nodedist, \invsqrttwo*0.5*\nodedist);
        \draw[line width=\lw] (VR2) -- ++(-0.5*\nodedist, \invsqrttwo*0.5*\nodedist);
        \draw[line width=\lw] (R) -- ++(-0.5*\nodedist, -1*\invsqrttwo*0.5*\nodedist);
        \draw[line width=\lw] (R) -- ++(0.5*\nodedist, -1*\invsqrttwo*0.5*\nodedist);
        \draw[line width=\lw, dashed] ($(VL2) + (0.5*\nodedist, \invsqrttwo*0.5*\nodedist)$) -- ($(R) + (-0.5*\nodedist, -1*\invsqrttwo*0.5*\nodedist)$);
        \draw[line width=\lw, dashed] ($(VR2) + (-0.5*\nodedist, \invsqrttwo*0.5*\nodedist)$) -- ($(R) + (0.5*\nodedist, -1*\invsqrttwo*0.5*\nodedist)$);
        \draw[line width=\lw, |-|] (4.3*\nodedist,0) -- ++(0,-4*\invsqrttwo*\nodedist)node[midway,right]{$L$};
    \end{scope}

    \begin{scope}[shift={(15*\nodedist,0)}]
        \foreach \y in {0,1,3,4}{
            \node[csimnode] (V\y) at (0,-1*\y*\nodedist){};
            \foreach \x in {1,3,4}{
                \node[csimnode] (Q\x\y) at (\x*\nodedist,-1*\y*\nodedist){};
                \draw[line width=\lw] (Q\x\y) -- ++(\invsqrttwo*\physdist,\invsqrttwo*\physdist);
            }
            \draw[line width=\lw] (V\y) -- (Q1\y) -- ++(0.5*\nodedist,0);
            \draw[line width=\lw, dashed] ($(Q1\y) + (0.5*\nodedist,0)$) -- ($(Q3\y) + (-0.5*\nodedist,0)$);
            \draw[line width=\lw] (Q4\y) -- (Q3\y) -- ++(-0.5*\nodedist,0);
        }
        \draw[line width=\lw] (V0) -- (V1) -- ++(0,-0.5*\nodedist);
        \draw[line width=\lw, dashed] ($(V1) + (0,-0.5*\nodedist)$) -- ($(V3) + (0,0.5*\nodedist)$);
        \draw[line width=\lw] (V4) -- (V3) -- ++(0,0.5*\nodedist);
        \draw[line width=\lw, |-|] ($(Q10) + (0,\nodedist)$) -- ($(Q40) + (0,\nodedist)$)node[midway,above] {$L$};
        \draw[line width=\lw, |-|] ($(Q40) + (\nodedist,0)$) -- ($(Q44) + (\nodedist,0)$)node[midway,right] {$L$};
    \end{scope}

    \draw[line width=\lw] (-4.5*\nodedist,2*\nodedist) rectangle (5*\nodedist,-5*\nodedist);
    \draw[line width=\lw] (5*\nodedist,2*\nodedist) rectangle (14*\nodedist,-5*\nodedist);
    \draw[line width=\lw] (14*\nodedist,2*\nodedist) rectangle (21*\nodedist,-5*\nodedist);
    % \node[anchor=north west] at (-4.5*\nodedist,2*\nodedist) {a)};
    % \node[anchor=north west] at (5*\nodedist,2*\nodedist) {b)};
    % \node[anchor=north west] at (14*\nodedist,2*\nodedist) {c)};
\end{tikzpicture}}

    % (a)
    \put(-113,85){(a)\phantomsubcaption\label{fig:rand_num_a}}

    % (b)
    \put(10,85){(b)\phantomsubcaption\label{fig:rand_num_b}}

    % (c)
    \put(127,85){(c)\phantomsubcaption\label{fig:rand_num_c}}
    \end{overpic}
    \caption{The three additional tree structures explored in ~\autoref{sec:rand_numerics}. a) T-Tree structure, b) binary tree structure, c) the fork tensor product structure. Here, only the TTNSs are depicted. The respective TTNOs have the exact same structure but with two physical legs per node.}
    \label{fig:rand_num_trees}
\end{figure*}

\subsection{Random States and Operators}\label{sec:rand_numerics}
\begin{figure*}
    \centering
    \begin{subfigure}{0.49\textwidth}
        \includegraphics{\figpath 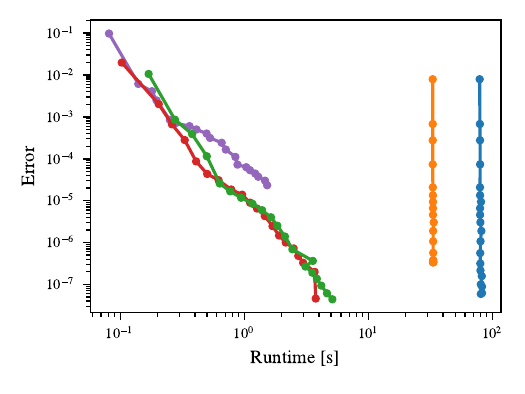}
        \caption{MPS}
    \end{subfigure}
    \begin{subfigure}{0.49\textwidth}
        \includegraphics{\figpath 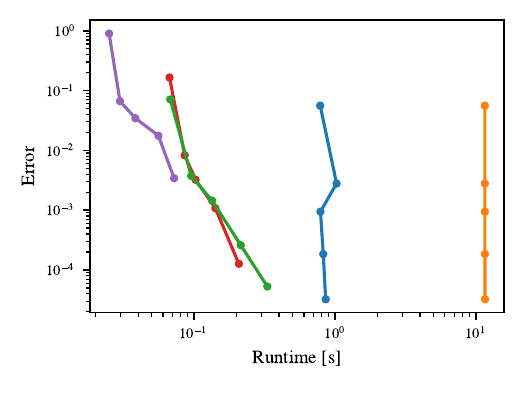}
        \caption{Binary TTNS}
    \end{subfigure}
    \begin{subfigure}{0.49\textwidth}
        \includegraphics{\figpath 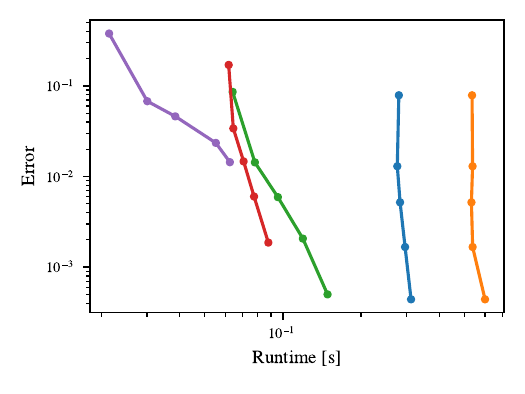}
        \caption{T-Tree TTNS}
    \end{subfigure}
    \begin{subfigure}{0.49\textwidth}
        \includegraphics{\figpath 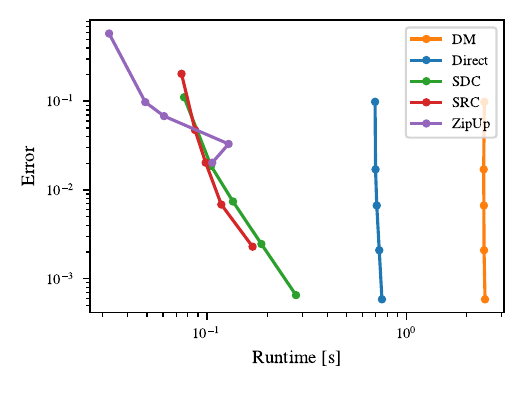}
        \caption{FTPS}
    \end{subfigure}
    \caption{Error versus runtime plots for the different tree structures, the results for our SDC method are shown in green. Each data point on a line corresponds to the results for one target bond dimension. In the case of the MPS, the bond dimensions range from $2$ to $40$, while for the other structures, they range from $2$ to $10$, increasing by $2$ with each consecutive data point in both cases.}
    \label{fig:random_num_plots}
\end{figure*}

As a first comparison of the methods, we applied random TTNOs to random TTNSs for varying desired bond dimensions in the regime $D_S=D_O=D$. Inspired by \cite{Camano2025}, both the real and imaginary parts of every element of the TTNOs and the TTNSs were drawn independently from a uniform random distribution on the interval $[-0.5,1]$. This choice is motivated by the fact that tensor network contraction of random tensor networks becomes easier if the tensor elements have a positive bias \cite{gray2024hyperoptimized, Jiang2025, Camano2025}. We conducted this experiment with four different types of trees. The first one is the tensor train structure with $50$ sites, an initial bond dimension of $D=40$ for both the MPS and the MPO, and a local dimension $d=3$. The bond dimension of the exactly contracted tensor networks is $40^2=1600$, though we only run our approximation methods until a maximum target bond dimension $\Bar{D}=40$, equal to the initial bond dimension.

The other three are T3NS with a restricted structure:
\begin{itemize}
    \item The \emph{T-Tree} structure, named due to its resemblance to the letter T. There is one central node connected to three chains of nodes with one physical leg each. In general, all three chains can have different numbers of nodes. However, we restrict our exploration to T-trees with the same length $L$ in each chain, allowing us to use $L$ as the defining system-size parameter. A T-tree is depicted in ~\autoref{fig:rand_num_a}. As a system size, we use $L=20$.
    \item The \emph{binary tree} structure. There is a predefined root node, and every node has at most two children. Additionally, only leaf nodes may have physical legs. This is one of the more commonly used tree structures \cite{Orus2019}. In our numerical experiment, we further restrict this structure by using only perfectly balanced binary trees. That is, every non-leaf node has exactly two children. Thus, we can use the tree's depth $L$ as the defining system size parameter. A binary TTNS is depicted in ~\autoref{fig:rand_num_b}. As a system size, we use $L=6$.
    \item The {fork tensor product (FTP)} structure \cite{Bauernfeind2017}. The FTP is a rectangular structure with a single chain of virtual nodes that lack a physical leg. Attached to each of these nodes is a chain of nodes with a physical leg. Again, we further restrict this structure, requiring that every chain has the same length $L$ and that the number of virtual nodes equals this length. Thus, we can use $L$ as the system-size parameter, defining a square shape. An FTPS is depicted in ~\autoref{fig:rand_num_c}. As a system size, we use $L=8$.
\end{itemize}

For all three structures, we can only run the exact contraction to a bond dimension of $D=10$, leading to a resulting bond dimension of $10^2=100$, but again with a local dimension $d=3$. This is due to the significantly higher memory requirement of the tensors with three virtual bonds.

The error sustained by the imperfect application of the TTNO to the TTNS is determined as
\begin{equation}
    \text{Err} = \left\| \ket{\psi'} - \hat{O} \ket{\psi} \right\|,
\end{equation}
where $\ket{\psi'}$ is the TTNS obtained via one of the application algorithms and $\hat{O} \ket{\psi}$ is the exact result obtained via the direct application without truncating the state afterwards.

The results of the compressions are collected in ~\autoref{fig:random_num_plots}. We first discuss the MPS case. We see that SDC and SRC perform almost identically and significantly better than the other methods. For a given bond dimension, the SDC tends to achieve a slightly lower error than the SRC. Both methods outperform the Zip-Up method in terms of accuracy: for a given error, a lower bond dimension is required, and especially for low errors and large bond dimensions, SDC and SRC have lower runtime. They also outperform the DM and direct methods in runtime by around one order of magnitude for a given error threshold, though the error is slightly worse for a given bond dimension. 

We next move to the three different T3NS structures. While the different initial bond dimensions make direct performance comparisons with the MPS case difficult, we observe that SDC and SRC perform similarly. Both SDC and SRC require similar runtimes to achieve a given error. For a given bond dimension, both achieve higher accuracy than the Zip-Up algorithm; for the small maximum bond dimensions considered here, up to the (relatively) low accuracy achieved by the Zip-Up algorithm, it requires slightly lower runtime in two out of the three tree structures, however, the slope of the Zip-Up error versus runtime curve suggests that at higher accuracies, SDC and SRC will be superior, similar to the MPS case. The other compression methods require, again, an order of magnitude more time, and compared to the MPS case, the DM compression method suddenly performs significantly worse. This is due to the significantly worse scaling compared to the other methods, as soon as a tensor with three virtual bonds is encountered, cf. ~\autoref{tab:time_scal} and ~\autoref{tab:mem_scal}.

\subsection{Circuit Simulation}

\begin{figure*}
    \begin{tikzpicture}[rotate=90]
    \def\sitedist{0.6}
    \def\layerdist{0.55}
    \foreach \i in {0,...,26}{
        \pgfmathsetmacro{\nexti}{int(\i + 1)}
        \node[anchor=north] at (0,-1*\i*\sitedist){$\nexti$};
        \draw[line width=\lw] (0,-1*\i*\sitedist) -- ++(10*\layerdist,0);
    }
    \foreach \cntrl/\other/\layer in {0/1/0, 2/3/0, 4/5/0, 6/7/0, 9/10/0, 11/12/0, 13/14/0, 15/16/0, 18/19/0, 20/21/0, 22/23/0, 24/25/0, 0/2/1, 4/6/1, 13/15/1, 18/20/1, 22/24/1, 1/3/2, 5/7/2, 10/12/2, 14/16/2, 19/21/2, 23/25/2, 0/2/3, 4/7/3, 9/12/3, 13/16/3, 18/21/3, 22/25/3, 1/2/4, 5/6/4, 10/11/4, 14/15/4, 19/20/4, 23/24/4, 3/8/5, 12/17/5, 21/26/5, 7/8/6, 16/17/6, 25/26/6, 8/17/7, 17/26/8, 8/26/9}{
        \node[copy] (A) at (0.5*\layerdist+\layer*\layerdist, -1*\sitedist*\cntrl) {};
        \node[copy] (B) at (0.5*\layerdist+\layer*\layerdist, -1*\sitedist*\other) {};
        \draw[line width=\lw] (A) -- (B);
    }

\end{tikzpicture}
    \caption{The structure of one batch of the quantum circuit. Two vertically connected dots represent a two-qubit gate.}
    \label{fig:quantum_circuit}
\end{figure*}
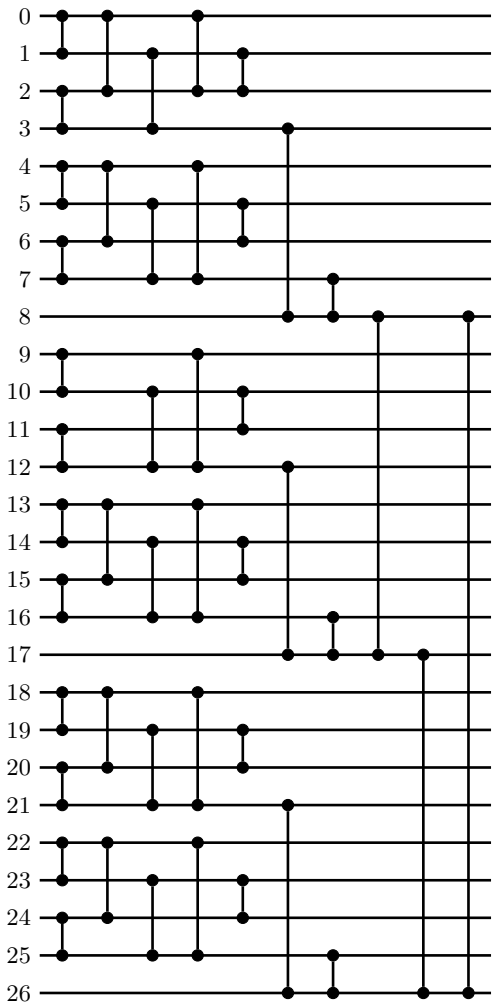

\begin{figure*}
    \begin{subfigure}{0.475\textwidth}
        \begin{tikzpicture}
    \def\nodedist{0.7}
    \def\textsize{\tiny}
    \def\physdist{0.5}
    \def\costhirty{0.866}
    \def\sinthirty{0.5}
    \def\cosfifteen{0.966}
    \def\sinfifteen{0.259}
    \node[csimnode] (R) at (0,0) {};
    \node[csimnode] (Q9) at ($(R) + (0,\nodedist)$) {{\textsize{9}}};
    \draw[line width=\lw] (Q9) -- ++(\physdist,0);
    \node[csimnode] (Vup) at ($(Q9) + (0,\nodedist)$) {};
    \foreach \i/\l in {1/4, 2/3, 3/2, 4/1}{
        \node[csimnode] (Q\l) at ($(Vup) + (-1*\i*\invsqrttwo*\nodedist, \i*\invsqrttwo*\nodedist)$) {{\textsize{\l}}};
        \draw[line width=\lw] (Q\l) -- ++(\invsqrttwo*\physdist, \invsqrttwo*\physdist);
    }
    \foreach \i/\l in {1/8, 2/7, 3/6, 4/5}{
        \node[csimnode] (Q\l) at ($(Vup) + (1*\i*\invsqrttwo*\nodedist, \i*\invsqrttwo*\nodedist)$) {{\textsize{\l}}};
        \draw[line width=\lw] (Q\l) -- ++(-1*\invsqrttwo*\physdist, \invsqrttwo*\physdist);
    }
    % Left part
    \node[csimnode] (Q18) at ($(R) + (-1*\costhirty*\nodedist,-1*\sinthirty*\nodedist)$) {{\textsize{18}}};
    \draw[line width=\lw] (Q18) -- ++(-\sinthirty*\physdist,\costhirty*\physdist);
    \node[csimnode] (Vleft) at ($(Q18) + (-1*\costhirty*\nodedist,-1*\sinthirty*\nodedist)$) {};
    \foreach \i/\l in {1/13, 2/12, 3/11, 4/10}{
        \node[csimnode] (Q\l) at ($(Vleft) + (-1*\i*\cosfifteen*\nodedist, \i*\sinfifteen*\nodedist)$) {{\textsize{\l}}};
        \draw[line width=\lw] (Q\l) -- ++(0, \physdist);
    }
    \foreach \i/\l in {1/17, 2/16, 3/15, 4/14}{
        \node[csimnode] (Q\l) at ($(Vleft) + (-1*\i*\sinfifteen*\nodedist, -1*\i*\cosfifteen*\nodedist)$) {{\textsize{\l}}};
        \draw[line width=\lw] (Q\l) -- ++(-1*\physdist,0);
    }
    % Right part
    \node[csimnode] (Q27) at ($(R) + (\costhirty*\nodedist,-1*\sinthirty*\nodedist)$) {{\textsize{27}}};
    \draw[line width=\lw] (Q27) -- ++(-\sinthirty*\physdist,-\costhirty*\physdist);
    \node[csimnode] (Vright) at ($(Q27) + (\costhirty*\nodedist,-1*\sinthirty*\nodedist)$) {};
    \foreach \i/\l in {1/26, 2/25, 3/24, 4/23}{
        \node[csimnode] (Q\l) at ($(Vright) + (\i*\cosfifteen*\nodedist, \i*\sinfifteen*\nodedist)$) {{\textsize{\l}}};
        \draw[line width=\lw] (Q\l) -- ++(0, \physdist);
    }
    \foreach \i/\l in {1/22, 2/21, 3/20, 4/19}{
        \node[csimnode] (Q\l) at ($(Vright) + (\i*\sinfifteen*\nodedist, -1*\i*\cosfifteen*\nodedist)$) {{\textsize{\l}}};
        \draw[line width=\lw] (Q\l) -- ++(\physdist,0);
    }
    % Lines
    \draw[line width=\lw] (R) -- (Q9) -- (Vup);
    \draw[line width=\lw] (R) -- (Q18) -- (Vleft);
    \draw[line width=\lw] (R) -- (Q27) -- (Vright);
    \draw[line width=\lw] (Q1) -- (Q2) -- (Q3) -- (Q4) -- (Vup) -- (Q8) -- (Q7) -- (Q6) -- (Q5);
    \draw[line width=\lw] (Q10) -- (Q11) -- (Q12) -- (Q13) -- (Vleft) -- (Q17) -- (Q16) -- (Q15) -- (Q14);
    \draw[line width=\lw] (Q19) -- (Q20) -- (Q21) -- (Q22) -- (Vright) -- (Q26) -- (Q25) -- (Q24) -- (Q23);
\end{tikzpicture}
        \caption{In this structure, groups of sites with high interaction are placed on chains, and only the site interacting with the remaining system is connected to the remaining tensor network. The three sites that facilitate interaction between the highly interacting regions are placed on intermediate tensors attached to the root.}
        \label{fig:csim_t3ns}
    \end{subfigure}
    \begin{subfigure}{0.475\textwidth}
        \begin{tikzpicture}
    \def\nodedist{0.7}
    \def\textsize{\tiny}
    \def\physdist{0.5}
    \def\costhirty{0.866}
    \def\sinthirty{0.5}
    \def\cosfifteen{0.966}
    \def\sinfifteen{0.259}
    \node[csimnode] (R) at (0,0) {};
    \node[csimnode] (V1) at ($(R) + (0,\nodedist)$) {};
    \node[csimnode] (Q8) at ($(V1) + (\nodedist,0)$) {\textsize{9}};
    \draw[line width=\lw] (Q8) -- ++(\physdist,0);
    \node[csimnode] (V2) at ($(V1) + (0,\nodedist)$) {};
    \node[csimnode] (V3) at ($(V2) + (-1*\invsqrttwo*\nodedist,\invsqrttwo*\nodedist)$) {};
    \node[csimnode] (Q3) at ($(V3) + (-1*\invsqrttwo*\nodedist,-1*\invsqrttwo*\nodedist)$) {\textsize{4}};
    \draw[line width=\lw] (Q3) -- ++(-1*\invsqrttwo*\physdist,-1*\invsqrttwo*\physdist);
    \node[csimnode] (V4) at ($(V3) + (-1*\invsqrttwo*\nodedist,\invsqrttwo*\nodedist)$) {};
    \node[csimnode] (Q2) at ($(V4) + (-1*\invsqrttwo*\nodedist,-1*\invsqrttwo*\nodedist)$) {\textsize{3}};
    \draw[line width=\lw] (Q2) -- ++(-1*\invsqrttwo*\physdist,-1*\invsqrttwo*\physdist);
    \node[csimnode] (Q1) at ($(V4) + (-1*\invsqrttwo*\nodedist,1*\invsqrttwo*\nodedist)$) {\textsize{2}};
    \draw[line width=\lw] (Q1) -- ++(-1*\invsqrttwo*\physdist,1*\invsqrttwo*\physdist);
    \node[csimnode] (Q0) at ($(V4) + (1*\invsqrttwo*\nodedist,1*\invsqrttwo*\nodedist)$) {\textsize{1}};
    \draw[line width=\lw] (Q0) -- ++(1*\invsqrttwo*\physdist,1*\invsqrttwo*\physdist);
    \draw[line width=\lw] (V2) -- (V3) -- (V4) -- (Q1);
    \draw[line width=\lw] (Q2) -- (V4) -- (Q0);
    \draw[line width=\lw] (Q3) -- (V3);
    \node[csimnode] (V5) at ($(V2) + (1*\invsqrttwo*\nodedist,\invsqrttwo*\nodedist)$) {};
    \node[csimnode] (Q7) at ($(V5) + (1*\invsqrttwo*\nodedist,-1*\invsqrttwo*\nodedist)$) {\textsize{8}};
    \draw[line width=\lw] (Q7) -- ++(1*\invsqrttwo*\physdist,-1*\invsqrttwo*\physdist);
    \node[csimnode] (V6) at ($(V5) + (1*\invsqrttwo*\nodedist,\invsqrttwo*\nodedist)$) {};
    \node[csimnode] (Q4) at ($(V6) + (1*\invsqrttwo*\nodedist,-1*\invsqrttwo*\nodedist)$) {\textsize{5}};
    \draw[line width=\lw] (Q4) -- ++(1*\invsqrttwo*\physdist,-1*\invsqrttwo*\physdist);
    \node[csimnode] (Q5) at ($(V6) + (1*\invsqrttwo*\nodedist,1*\invsqrttwo*\nodedist)$) {\textsize{6}};
    \draw[line width=\lw] (Q5) -- ++(1*\invsqrttwo*\physdist,1*\invsqrttwo*\physdist);
    \node[csimnode] (Q6) at ($(V6) + (-1*\invsqrttwo*\nodedist,1*\invsqrttwo*\nodedist)$) {\textsize{7}};
    \draw[line width=\lw] (Q6) -- ++(-1*\invsqrttwo*\physdist,1*\invsqrttwo*\physdist);
    \draw[line width=\lw] (V2) -- (V5) -- (V6) -- (Q5);
    \draw[line width=\lw] (Q6) -- (V6) -- (Q4);
    \draw[line width=\lw] (Q7) -- (V5);
    \draw[line width=\lw] (R) -- (V1) -- (V2);
    \draw[line width=\lw] (V1) -- (Q8);
    % Left part
    \node[csimnode] (V7) at ($(R) + (-1*\costhirty*\nodedist,-1*\sinthirty*\nodedist)$) {};
    \node[csimnode] (Q17) at ($(V7) + (-1*\sinthirty*\nodedist,1*\costhirty*\nodedist)$) {{\textsize{18}}};
    \draw[line width=\lw] (Q17) -- +(-1*\sinthirty*\physdist,1*\costhirty*\physdist);
    \node[csimnode] (V8) at ($(V7) + (-1*\costhirty*\nodedist,-1*\sinthirty*\nodedist)$) {};
    \node[csimnode] (V9) at ($(V8) + (-1*\cosfifteen*\nodedist, \sinfifteen*\nodedist)$) {};
    \node[csimnode] (Q12) at ($(V9) + (1*\sinfifteen*\nodedist,1*\cosfifteen*\nodedist)$) {{\textsize{13}}};
    \draw[line width=\lw] (Q12) -- +(1*\sinfifteen*\physdist,1*\cosfifteen*\physdist);
    \node[csimnode] (V10) at ($(V9) + (-1*\cosfifteen*\nodedist, \sinfifteen*\nodedist)$) {};
    \node[csimnode] (Q11) at ($(V10) + (1*\sinfifteen*\nodedist,1*\cosfifteen*\nodedist)$) {{\textsize{12}}};
    \draw[line width=\lw] (Q11) -- +(1*\sinfifteen*\physdist,1*\cosfifteen*\physdist);
    \node[csimnode] (Q10) at ($(V10) + (-1*\cosfifteen*\nodedist, \sinfifteen*\nodedist)$) {{\textsize{11}}};
    \draw[line width=\lw] (Q10) -- +(-1*\cosfifteen*\physdist,1*\sinfifteen*\physdist);
    \node[csimnode] (Q9) at ($(V10) + (-1*\sinfifteen*\nodedist,-1*\cosfifteen*\nodedist)$) {{\textsize{10}}};
    \draw[line width=\lw] (Q9) -- +(-1*\sinfifteen*\physdist,-1*\cosfifteen*\physdist);
    \draw[line width=\lw] (V8) -- (V9) -- (V10) -- (Q10);
    \draw[line width=\lw] (Q9) -- (V10) -- (Q11);
    \draw[line width=\lw] (Q12) -- (V9);
    \node[csimnode] (V11) at ($(V8) + (-1*\sinfifteen*\nodedist, -1*\cosfifteen*\nodedist)$) {};
    \node[csimnode] (Q16) at ($(V11) + (1*\cosfifteen*\nodedist, -1*\sinfifteen*\nodedist)$) {{\textsize{17}}};
    \draw[line width=\lw] (Q16) -- +(1*\cosfifteen*\physdist,-1*\sinfifteen*\physdist);
    \node[csimnode] (V12) at ($(V11) + (-1*\sinfifteen*\nodedist, -1*\cosfifteen*\nodedist)$) {};
    \node[csimnode] (Q13) at ($(V12) + (-1*\cosfifteen*\nodedist, 1*\sinfifteen*\nodedist)$) {{\textsize{14}}};
    \draw[line width=\lw] (Q13) -- +(-1*\cosfifteen*\physdist,1*\sinfifteen*\physdist);
    \node[csimnode] (Q14) at ($(V12) + (-1*\sinfifteen*\nodedist, -1*\cosfifteen*\nodedist)$) {{\textsize{15}}};
    \draw[line width=\lw] (Q14) -- +(-1*\sinfifteen*\physdist,-1*\cosfifteen*\physdist);
    \node[csimnode] (Q15) at ($(V12) + (1*\cosfifteen*\nodedist, -1*\sinfifteen*\nodedist)$) {{\textsize{16}}};
    \draw[line width=\lw] (Q15) -- +(1*\cosfifteen*\physdist,-1*\sinfifteen*\physdist);
    \draw[line width=\lw] (V8) -- (V11) -- (V12) -- (Q14);
    \draw[line width=\lw] (Q13) -- (V12) -- (Q15);
    \draw[line width=\lw] (Q16) -- (V11);
    \draw[line width=\lw] (Q17) -- (V7);
    \draw[line width=\lw] (V8) -- (V7) -- (R);
    % Right part
    \node[csimnode] (V13) at ($(R) + (1*\costhirty*\nodedist,-1*\sinthirty*\nodedist)$) {};
    \node[csimnode] (Q26) at ($(V13) + (-1*\sinthirty*\nodedist,-1*\costhirty*\nodedist)$) {{\textsize{27}}};
    \draw[line width=\lw] (Q26) -- +(-1*\sinthirty*\physdist,-1*\costhirty*\physdist);
    \node[csimnode] (V14) at ($(V13) + (1*\costhirty*\nodedist,-1*\sinthirty*\nodedist)$) {};
    \node[csimnode] (V15) at ($(V14) + (1*\cosfifteen*\nodedist, \sinfifteen*\nodedist)$) {};
   \node[csimnode] (Q25) at ($(V15) + (-1*\sinfifteen*\nodedist,1*\cosfifteen*\nodedist)$) {{\textsize{26}}};
    \draw[line width=\lw] (Q25) -- +(-1*\sinfifteen*\physdist,1*\cosfifteen*\physdist);
    \node[csimnode] (V16) at ($(V15) + (1*\cosfifteen*\nodedist, \sinfifteen*\nodedist)$) {};
    \node[csimnode] (Q24) at ($(V16) + (-1*\sinfifteen*\nodedist,1*\cosfifteen*\nodedist)$) {{\textsize{25}}};
    \draw[line width=\lw] (Q24) -- +(-1*\sinfifteen*\physdist,1*\cosfifteen*\physdist);
    \node[csimnode] (Q23) at ($(V16) + (1*\cosfifteen*\nodedist, \sinfifteen*\nodedist)$) {{\textsize{24}}};
    \draw[line width=\lw] (Q23) -- +(1*\cosfifteen*\physdist,1*\sinfifteen*\physdist);
    \node[csimnode] (Q22) at ($(V16) + (1*\sinfifteen*\nodedist,-1*\cosfifteen*\nodedist)$) {{\textsize{23}}};
    \draw[line width=\lw] (Q22) -- +(1*\sinfifteen*\physdist,-1*\cosfifteen*\physdist);
    \draw[line width=\lw] (V14) -- (V15) -- (V16) -- (Q23);
    \draw[line width=\lw] (Q22) -- (V16) -- (Q24);
    \draw[line width=\lw] (Q25) -- (V15);
    \node[csimnode] (V17) at ($(V14) + (1*\sinfifteen*\nodedist, -1*\cosfifteen*\nodedist)$) {};
    \node[csimnode] (Q21) at ($(V17) + (-1*\cosfifteen*\nodedist, -1*\sinfifteen*\nodedist)$) {{\textsize{22}}};
    \draw[line width=\lw] (Q21) -- +(-1*\cosfifteen*\physdist,-1*\sinfifteen*\physdist);
    \node[csimnode] (V18) at ($(V17) + (1*\sinfifteen*\nodedist, -1*\cosfifteen*\nodedist)$) {};
    \node[csimnode] (Q18) at ($(V18) + (1*\cosfifteen*\nodedist, 1*\sinfifteen*\nodedist)$) {{\textsize{19}}};
    \draw[line width=\lw] (Q18) -- +(1*\cosfifteen*\physdist,1*\sinfifteen*\physdist);
    \node[csimnode] (Q19) at ($(V18) + (1*\sinfifteen*\nodedist, -1*\cosfifteen*\nodedist)$) {{\textsize{20}}};
    \draw[line width=\lw] (Q19) -- +(1*\sinfifteen*\physdist,-1*\cosfifteen*\physdist);
    \node[csimnode] (Q20) at ($(V18) + (-1*\cosfifteen*\nodedist, -1*\sinfifteen*\nodedist)$) {{\textsize{21}}};
    \draw[line width=\lw] (Q20) -- +(-1*\cosfifteen*\physdist,-1*\sinfifteen*\physdist);
    \draw[line width=\lw] (V14) -- (V17) -- (V18) -- (Q19);
    \draw[line width=\lw] (Q20) -- (V18) -- (Q18);
    \draw[line width=\lw] (Q21) -- (V17);
    \draw[line width=\lw] (Q26) -- (V13);
    \draw[line width=\lw] (V14) -- (V13) -- (R);
\end{tikzpicture}
        \caption{In this structure, groups of sites that interact strongly are represented by tensors that are close to each other. However, the tensors are not directly connected. On the other hand, the three sites that facilitate interaction between these groups are not directly attached to the root but are represented by a leaf tensor.}
        \label{fig:csim_binary}
    \end{subfigure}
    \caption{The two different T3NS structures used in the circuit simulation.}
\end{figure*}
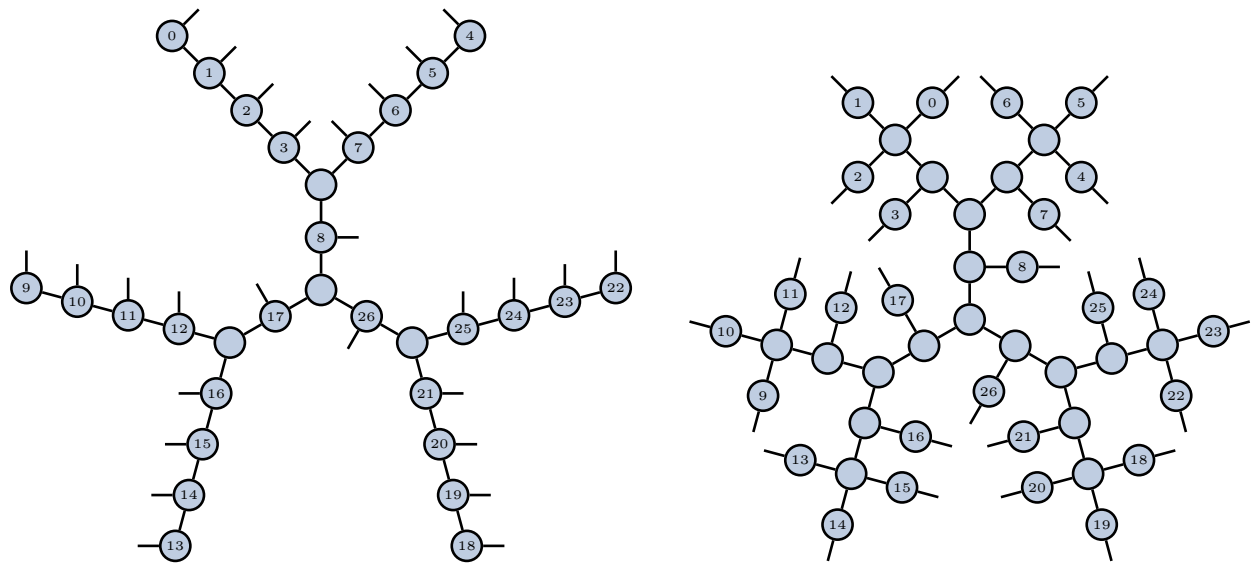

As a second experiment, we conduct a simulation of a quantum circuit $U_{\text{full}}$. To avoid the need for a reference calculation, we consider a quantum circuit of the form
\begin{equation}
    U_{\text{full}} = U U^\dagger.
\end{equation}
Accordingly, the output state should be equal to the input state for an exact circuit simulation. The circuit $U$ itself consists of $N$ batches
\begin{equation}
    U = \prod_{i=1}^N U_i,
\end{equation}
each batch $U_i$ has the form shown in ~\autoref{fig:quantum_circuit} and is thus a tree-like circuit \cite{Seitz2023}. In the circuit, every two-qubit gate $G$ is of the form
\begin{equation}
    G = \text{CNOT} \cdot (V_1 \otimes V_2),
\end{equation}
where all the single-qubit gates $V_i$ are drawn Haar-randomly. Note that the $V_i$ are only placed in front of the CNOT, as they do not have any entangling power. Thus, multiple $V_i$ gates applied one after another do not impact the performance of the tensor network method compared to a single such gate. We construct every level of the circuit as shown in ~\autoref{fig:quantum_circuit} as a separate TTNO, via the method introduced in \cite{Cakir2025}. The resulting TTNO is then applied to the current TTNS. We run this experiment on three different tree structures. The first is a simple 27-site MPS that is agnostic to the circuit's structure. The other two structures are more complex trees, specifically adapted to the entangling structure of the batches $U_i$. Both structures are T3NS, but one has the highly entangled gates combined into MPS-like structures, shown in ~\autoref{fig:csim_t3ns}. In the other T3NS, only leaf nodes have physical legs.

\begin{figure*}
    \begin{subfigure}{\textwidth}
        \includegraphics{\figpath 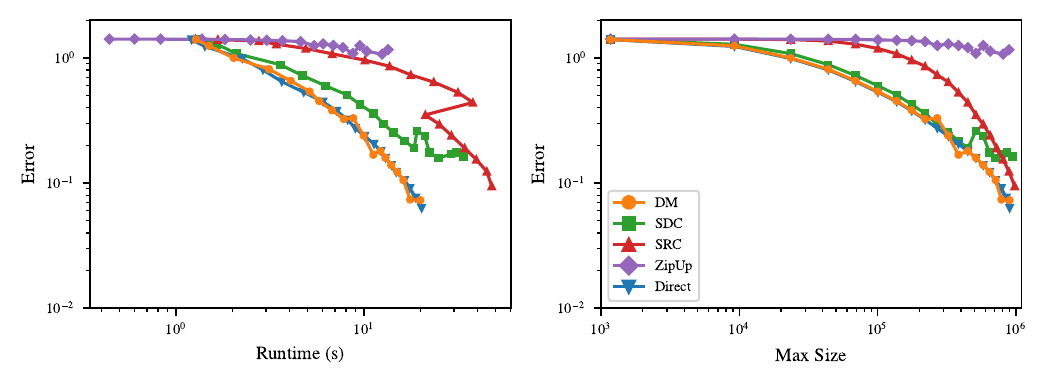}
        \caption{Results for the MPS}
    \end{subfigure}
    \begin{subfigure}{\textwidth}
        \includegraphics{\figpath 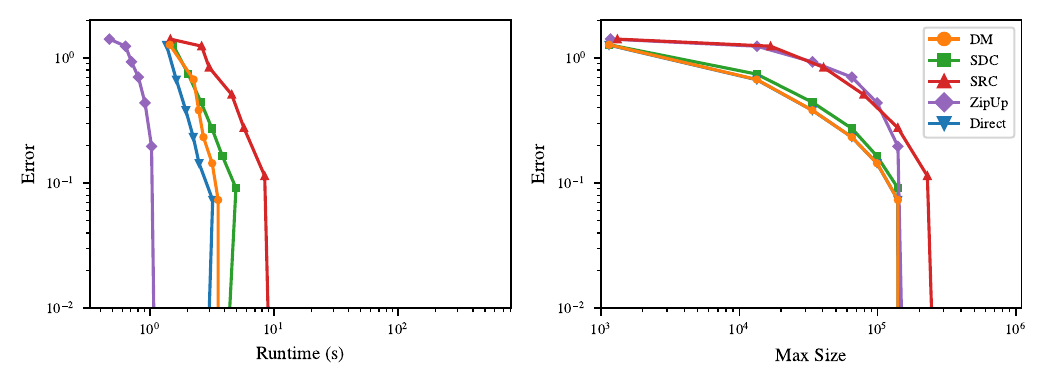}
        \caption{Results for the T3NS with MPS chains as subtree, depicted in ~\autoref{fig:csim_t3ns}.}
    \end{subfigure}
    \begin{subfigure}{\textwidth}
        \includegraphics{\figpath 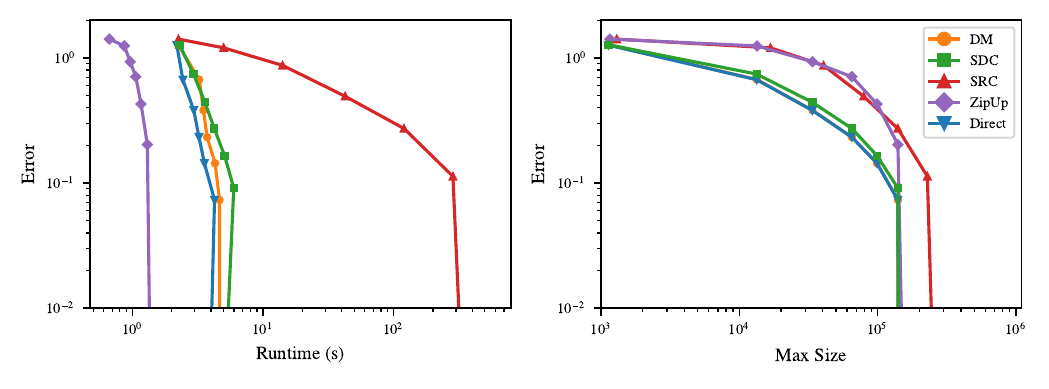}
        \caption{Results for the T3NS where only leafs have physical legs, depicted in ~\autoref{fig:csim_binary}.}
    \end{subfigure}
    \caption{Runtime (left) and maximum tensor network size during the run (right) plotted against the error of the performed circuit simulation for three different tree structures. Each point corresponds to a different maximally allowed bond dimension.}
    \label{fig:circ_sim_results}
\end{figure*}

We ran this simulation for all tree structures with $N=3$ circuit layers and bond dimensions $\Bar{D}$ ranging from $10$ to $200$. The quantum system starts in the computational all-zero state $\ket{\psi_0} = \ket{0}$, and we record the error when comparing the final to the initial state
\begin{equation}
    \text{Err} = \left\| \ket{\psi_0} - U_{\text{full}} \ket{\psi_0} \right\|.
\end{equation}
To compare the performance of the different methods, we not only recorded the runtime but also the maximum size, i.e., the maximum number of tensor entries in the TTNS, during runtime, as a measure of the memory requirement. The results averaged over five runs with different seeds for the random gates are plotted in ~\autoref{fig:circ_sim_results}. 

The MPS simulations once again show that the SDC compression performs reasonably well and significantly outperforms SRC and Zip-Up. However, it is slightly outperformed by the DM and Direct methods. This can be attributed to two factors: first, the operator bond dimension $D_O$ of each circuit level is small, ranging from $1$ to $4$, which is very favorable for the Direct and DM methods, and second, the calculations are performed in the very low-accuracy regime because the circuit is highly incompressible, which means that absolute compression accuracy is the most important factor. If we consider the number of tensor elements in ~\autoref{fig:circ_sim_results}, we see that in fact Direct, DM, and SDC achieve the most compact compression out of all the methods, and this suggests that for slightly larger $D_O$, SDC will once again outperform the Direct and DM approaches. In contrast, ZipUp, which has some of the best asymptotic scaling, has the lowest accuracy and, in fact, cannot reduce the error as the bond dimension increases. 

For the two tree structures, all methods converge to numerical error at a bond dimension of $\Bar{D} = 35$, indicating the limited entanglement generated within these structures. However, the MPS simulations cannot achieve an error below $10^{-2}$ at all, even for a bond dimension $\Bar{D}=100$. Notably, the TTNS also requires similar or fewer resources, both in terms of memory and runtime, compared to the MPS to achieve the same error. This happens because the T3NSs resulting from the application of a circuit-level are far more compressible than the MPS. This supports that long-range interactions cause problems for less optimised structures, thereby demonstrating that TTN can provide an advantage over the simpler tensor train structure. When examining the performance of the different methods, we note that Zip-Up is the fastest for tree structures. This is consistent with the highly compressible nature of the state here; i.e., the quality of the compression is less important, which, in this case, allows Zip-Up's low scaling and prefactor to outweigh the improved compression of the other methods. SRC is significantly slower than the SDC, DM and direct application methods. This is the most significant for the tree with only leaf phyiscal legs~\autoref{fig:csim_binary}, which has the most tensors with more than two virtual legs. For the other three methods (Direct, DM, SDC), the costs are quite similar. The Direct, DM, and SDC methods perform practically the same when considering the number of tensor elements required to achieve a given error, indicating that they achieve a similar level of compression in this case.

\section{Discussion}

We have introduced a new tensor network compression method, stochastic deterministic compression (SDC), in the context of applying a tree-tensor network operator to a tree-tensor network state, with the common application of a matrix product operator to a matrix product state being a special case.  Under reasonable assumptions, asymptotic analysis suggests that stochastic deterministic compression should be amongst the most performant of tensor network compression schemes, although the many compounding factors (tensor network operator dimension, desired accuracy, physical bond dimension, etc.) mean that real-world performance will depend on details of the compression problem at hand.  We have tested this compression technique in two settings: a random state generated by applying random tree tensor network operators, and a structured circuit with standard unitary entangling gates. We find that under most settings, stochastic deterministic compression is comparable to or outperforms the recently introduced stochastic randomized compression. With respect to other techniques, the comparison is more nuanced, but for problems where high-quality compression is needed (i.e. the tensor network state is not very compressible), it is either the best or amongst the best-performing techniques. Additionally, our results support the utility of more complex TTN structures tailored to the entanglement of the problem at hand, as a way to further improve classical methods for simulating quantum systems. Overall, we find stochastic deterministic compression to be a useful new tool in the arsenal of tensor network compression techniques, and one which we can recommend as a primitive in a wide range of tensor network simulations. 

% current state-of-the-art methods, such as SRC and DM. We also found that the choice of method can be significantly affected by the relative bond dimension of TTNO to TTNS. In all our numerical studies, the SDC method consistently performed among the best. Another advantage of our method over the SRC is the simpler implementation of an automatically adaptive bond dimension by setting a tolerance for the SVD. The flip side is that the SRC only requires contractions and QR decompositions, which yield a much greater performance boost when running simulations on GPUs \cite{Unfried2023}. Additionally, our results support the utility of more complex TTN structures tailored to the problem at hand, further improving the classical methods for simulating quantum systems. We can also conclude that the Zip-Up method tends to perform worse than the SDC and SRC methods when a desired error threshold is considered. Additionally, we found that the DM-based compression method, while useful for MPS, is unsuited for general T3NS due to significantly worse scaling compared to MPS structures. Notably, this gap can be filled by the SRC and SDC methods.

\section*{Acknowledgments}
The research is part of the Munich Quantum Valley, which is supported by the Bavarian state government with funds from the Hightech Agenda Bayern Plus. The visit of RMM to the California Institute of Technology, during which part of the research for this work was carried out, was additionally supported by the Internationalisierungsförderung of the TUM Graduate School. SS acknowledges funding by the BMW Group. Research at Caltech was supported by the US Department of Energy, Office of Science, via grant no. DE-SC0019374.\\

\section*{Author Contributions Statement}
GKLC conceived the project. RMM developed the method, implemented it, and performed the simulations. JG and CBM contributed to the development of the method. SS contributed to the implementation of the method. All authors discussed the results and contributed to the final manuscript.

\section*{Competing Interests}
The authors declare no competing interests.

\section*{Data Availability}
The code to run the simulations is available within the PyTreeNet library \cite{PyTreeNet, Milbradt2024B}, while the generated data can be found at \cite{Data}.

\bibliography{references}

\begin{thebibliography}{73}
\bibitem [{DMC()}]{DMCNote}%
  \BibitemOpen
  \href@noop {} {}\bibinfo {note} {The DMC algorithm has no first paper associated with it. However, it is included in the Itensor tensor network library~\cite{ItensorPub, ItensorCode} and pedagogically explained in~\cite{tensornetworkorg}.}\BibitemShut {Stop}%
\end{thebibliography}

\appendix
\section{Illustration of SDC for an MPS}\label{app:mps_illustration}
This appendix illustrates the various steps of the SDC for a length-$4$ MPS. The first step is the construction of the left subtree tensors $\mathcal{L}$ via repeated contraction and truncation until we reach the right-most site.

\begin{widetext}
\begin{equation}
\raisebox{-1cm}{
    \begin{tikzpicture}
    \foreach \i in {0,1,2,3}{
        \node[statenode] (S\i) at (\i*\mpsvirtdist,0){};
        \node[operatornode] (O\i) at (\i*\mpsvirtdist,\mpsinnerphysdist){};
        \draw[line width=\lw] (S\i) -- (O\i) -- ++(0,\mpsouterphysdist);
    }
    \draw[line width=\lw] (S0) -- (S1) -- (S2) -- (S3);
    \draw[line width=\lw] (O0) -- (O1) -- (O2) -- (O3);

    \node at (3.5*\mpsvirtdist,0.5*\mpsinnerphysdist){$\rightarrow$};

    \begin{scope}[shift={(4*\mpsvirtdist,0)}]
        \node (S0) at (0,0){};
        \node (O0) at (0,\mpsinnerphysdist){};
        \foreach \i in {1,2,3}{
            \node[statenode] (S\i) at (\i*\mpsvirtdist,0){};
            \node[operatornode] (O\i) at (\i*\mpsvirtdist,\mpsinnerphysdist){};
            \draw[line width=\lw] (S\i) -- (O\i) -- ++(0,\mpsouterphysdist);
        }
        \draw[line width=\lw] (S0) -- (S1) -- (S2) -- (S3);
        \draw[line width=\lw] (O0) -- (O1) -- (O2) -- (O3);
        \node[submps] (L) at (0,0.5*\mpsinnerphysdist){};
        \draw[line width=\lw] (L.north) -- ($(O1) + (-1*\mpsvirtdist,\mpsouterphysdist)$);
    \end{scope}

    \node at (7.5*\mpsvirtdist,0.5*\mpsinnerphysdist){$\rightarrow$};

    \begin{scope}[shift={(8*\mpsvirtdist,0)}]
        \node (S0) at (0,0){};
        \node (O0) at (0,\mpsinnerphysdist){};
        \foreach \i in {1}{
            \node[statenode] (S\i) at (\i*\mpsvirtdist,0){};
            \node[operatornode] (O\i) at (\i*\mpsvirtdist,\mpsinnerphysdist){};
            \draw[line width=\lw] (S\i) -- (O\i) -- ++(0,\mpsouterphysdist);
        }
        \draw[line width=\lw] (S0) -- (S1);
        \draw[line width=\lw] (O0) -- (O1);
        \node[submps] (L) at (0,0.5*\mpsinnerphysdist){};
        \draw[line width=\lw] (L.north) -- ($(O1) + (-1*\mpsvirtdist,\mpsouterphysdist)$);
    \end{scope}

    \node at (9.5*\mpsvirtdist,0.5*\mpsinnerphysdist){$=$};

    \begin{scope}[shift={(10.5*\mpsvirtdist,0)}]
        \node[statenode, fill=morange!40] (S) at (0,0.5*\mpsinnerphysdist){};
        \draw[line width=\lw] (S) -- ++(-0.5*\mpsvirtdist,0) -- ++(0,\mpsouterphysdist);
        \draw[line width=\lw] (S) -- ++(0,\mpsouterphysdist);
        \draw[tensorcut] ($(S) + (-0.4*\mpsvirtdist,0.4*\mpsouterphysdist)$) -- ($(S) + (0.4*\mpsvirtdist,-0.4*\mpsouterphysdist)$);
    \end{scope}

    \node at (11*\mpsvirtdist,0.5*\mpsinnerphysdist){$=$};

    \begin{scope}[shift={(13*\mpsvirtdist,0.5*\mpsinnerphysdist)}]
        \node[isonode, rotate=180, fill=morange!40] (S) at (0,0){};
        \node[matrix, fill=myellow!40] (M) at (-1*\mpsvirtdist,0){};
        \draw[line width=\lw] (S.north) to[out=0,in=-90] ($(S)+(0,\mpsouterphysdist)$);
        \draw[line width=\lw] (S) -- (M) -- ++(-1*\mpsouterphysdist,0);
    \end{scope}
\end{tikzpicture}
    }.
\end{equation}
\end{widetext}

In the end, we obtain a tensor of dimensions $\Bar{D}$ and $d$, shown in orange in the previous equation. It is decomposed via the QR decomposition to yield an isometric tensor, which will be the rightmost tensor in the new MPS, while the non-isometric matrix is no longer needed. We then proceed to one site to the left. Here, we reuse the left subtree tensor obtained in the previous step, and we project the right-most site to the desired bond dimension, yielding

\begin{widetext}
\begin{equation}
\raisebox{-1cm}{
    \begin{tikzpicture}
    \node (S0) at (0,0){};
    \node (O0) at (0,\mpsinnerphysdist){};
    \foreach \i in {1,2}{
        \node[statenode] (S\i) at (\i*\mpsvirtdist,0){};
        \node[operatornode] (O\i) at (\i*\mpsvirtdist,\mpsinnerphysdist){};
        \draw[line width=\lw] (S\i) -- (O\i) -- ++(0,\mpsouterphysdist);
    }
    \draw[line width=\lw] (S0) -- (S1) -- (S2);
    \draw[line width=\lw] (O0) -- (O1) -- (O2);
    \node[submps] (L) at (0,0.5*\mpsinnerphysdist){};
    \draw[line width=\lw] (L.north) -- ($(O1) + (-1*\mpsvirtdist,\mpsouterphysdist)$);
    \node[isonode, fill=morange!40, rotate=90] (Qs) at ($(O2) + (0,\mpsouterphysdist)$){{$*$}};
    \draw[line width=\lw] (Qs.south) -- ++(0,0.3*\mpsouterphysdist);

    \node at (2.5*\mpsvirtdist,0.5*\mpsinnerphysdist){$=$};
    
    \begin{scope}[shift={(3.5*\mpsvirtdist,0.5*\mpsinnerphysdist)}]
        \node[statenode, fill=morange!40] (S) at (0,0){};
        \draw[line width=\lw] (S) -- ++(-0.5*\mpsvirtdist,0) -- ++(0,\mpsouterphysdist);
        \draw[line width=\lw] (S) -- ++(0.5*\mpsvirtdist,0) -- ++(0,\mpsouterphysdist);
        \draw[line width=\lw] (S) -- ++(0,\mpsouterphysdist);
        \draw[tensorcut] ($(S) + (-0.4*\mpsvirtdist,0.4*\mpsouterphysdist)$) -- ($(S) + (0.4*\mpsvirtdist,-0.4*\mpsouterphysdist)$);
    \end{scope}

    \node at (4.5*\mpsvirtdist,0.5*\mpsinnerphysdist){$=$};

    \begin{scope}[shift={(6.5*\mpsvirtdist,0.5*\mpsinnerphysdist)}]
        \node[isonode, rotate=180, fill=morange!40] (S) at (0,0){};
        \node[matrix, fill=myellow!40] (M) at (-1*\mpsvirtdist,0){};
        \draw[line width=\lw] (S.north) to[out=0,in=-90] ($(S)+(0,\mpsouterphysdist)$);
        \draw[line width=\lw] (S) -- (M) -- ++(-1*\mpsouterphysdist,0);
        \draw[line width=\lw] (S) -- ++(0,-1*\mpsouterphysdist);
    \end{scope}
\end{tikzpicture}
    }.
\end{equation}
\end{widetext}

Where once more the final isometric tensor is the site's tensor in the new MPS. Moving to the second-to-last site, we again proceed similarly

\begin{widetext}
\begin{equation}
\raisebox{-1cm}{
    \begin{tikzpicture}
    \node (S0) at (0,0){};
    \node (O0) at (0,\mpsinnerphysdist){};
    \foreach \i in {1,2,3}{
        \node[statenode] (S\i) at (\i*\mpsvirtdist,0){};
        \node[operatornode] (O\i) at (\i*\mpsvirtdist,\mpsinnerphysdist){};
        \draw[line width=\lw] (S\i) -- (O\i) -- ++(0,\mpsouterphysdist);
    }
    \draw[line width=\lw] (S0) -- (S1) -- (S2) -- (S3);
    \draw[line width=\lw] (O0) -- (O1) -- (O2) -- (O3);
    \node[submps] (L) at (0,0.5*\mpsinnerphysdist){};
    \draw[line width=\lw] (L.north) -- ($(O1) + (-1*\mpsvirtdist,\mpsouterphysdist)$);
    \node[isonode, fill=morange!40, rotate=90] (Qs) at ($(O2) + (0,\mpsouterphysdist)$){{$*$}};
    \draw[line width=\lw] (Qs.south) -- ++(0,0.3*\mpsouterphysdist);
    \node[isonode, fill=morange!40, rotate=90] (Qs2) at ($(O3) + (0,\mpsouterphysdist)$){{$*$}};
    \draw[line width=\lw] (Qs2.south) to[out=90,in=0] (Qs.west);

    \node at (3.5*\mpsvirtdist,0.5*\mpsinnerphysdist){$=$};
    
    \begin{scope}[shift={(4.5*\mpsvirtdist,0.5*\mpsinnerphysdist)}]
        \node[statenode, fill=morange!40] (S) at (0,0){};
        \draw[line width=\lw] (S) -- ++(-0.5*\mpsvirtdist,0) -- ++(0,\mpsouterphysdist);
        \draw[line width=\lw] (S) -- ++(0.5*\mpsvirtdist,0) -- ++(0,\mpsouterphysdist);
        \draw[line width=\lw] (S) -- ++(0,\mpsouterphysdist);
        \draw[tensorcut] ($(S) + (-0.4*\mpsvirtdist,0.4*\mpsouterphysdist)$) -- ($(S) + (0.4*\mpsvirtdist,-0.4*\mpsouterphysdist)$);
    \end{scope}

    \node at (5.5*\mpsvirtdist,0.5*\mpsinnerphysdist){$=$};

    \begin{scope}[shift={(7.5*\mpsvirtdist,0.5*\mpsinnerphysdist)}]
        \node[isonode, rotate=180, fill=morange!40] (S) at (0,0){};
        \node[matrix, fill=myellow!40] (M) at (-1*\mpsvirtdist,0){};
        \draw[line width=\lw] (S.north) to[out=0,in=-90] ($(S)+(0,\mpsouterphysdist)$);
        \draw[line width=\lw] (S) -- (M) -- ++(-1*\mpsouterphysdist,0);
        \draw[line width=\lw] (S) -- ++(0,-1*\mpsouterphysdist);
    \end{scope}
\end{tikzpicture}
    }.
\end{equation}
\end{widetext}

Note that here the projection tensor of the rightmost site is contracted with the next projection tensor. Here, it is reasonable to reuse a partial contraction result from the previous site to avoid unnecessary computations. This reused part is exactly what we defined as the right subtree tensors earlier. If the MPS were longer, the above steps would be repeated until the last site is reached. For the last site, we merely contract

\begin{equation}
\raisebox{-1cm}{
    \begin{tikzpicture}
    \foreach \i in {0,1,2,3}{
        \node[statenode] (S\i) at (\i*\mpsvirtdist,0){};
        \node[operatornode] (O\i) at (\i*\mpsvirtdist,\mpsinnerphysdist){};
        \draw[line width=\lw] (S\i) -- (O\i) -- ++(0,\mpsouterphysdist);
    }
    \draw[line width=\lw] (S0) -- (S1) -- (S2) -- (S3);
    \draw[line width=\lw] (O0) -- (O1) -- (O2) -- (O3);
    \node[isonode, fill=morange!40, rotate=90] (Qs1) at ($(O1) + (0,\mpsouterphysdist)$){{$*$}};
    \draw[line width=\lw] (Qs1.south) -- ++(0,0.3*\mpsouterphysdist);
    \node[isonode, fill=morange!40, rotate=90] (Qs) at ($(O2) + (0,\mpsouterphysdist)$){{$*$}};
    \draw[line width=\lw] (Qs.south) to[out=90,in=0] (Qs1.west);
    \node[isonode, fill=morange!40, rotate=90] (Qs2) at ($(O3) + (0,\mpsouterphysdist)$){{$*$}};
    \draw[line width=\lw] (Qs2.south) to[out=90,in=0] (Qs.west);

    \node at (3.5*\mpsvirtdist,0.5*\mpsinnerphysdist){$=$};
    
    \begin{scope}[shift={(4.5*\mpsvirtdist,0.5*\mpsinnerphysdist)}]
        \node[statenode, fill=morange!40] (S) at (0,0){};
        \draw[line width=\lw] (S) -- ++(0.5*\mpsvirtdist,0) -- ++(0,\mpsouterphysdist);
        \draw[line width=\lw] (S) -- ++(0,\mpsouterphysdist);
    \end{scope}
\end{tikzpicture}
    }
\end{equation}

to obtain the tensor of the leftmost site. In total, the MPS resulting from the application of the MPO will then have the form

\begin{equation}
\hat{O}\ket{\psi} \approx
\raisebox{-0.5cm}{
    \begin{tikzpicture}
    \node[statenode, fill=morange!40] (S) at (0,0) {};
    \draw[line width=\lw] (S) -- ++(0,\mpsouterphysdist);
    \foreach \i in {1,2,3}{
        \node[isonode, rotate=180, fill=morange!40] (S\i) at (\i*\mpsvirtdist,0){};
        \draw[line width=\lw] (S\i.north) to[out=0,in=-90] ++(0,\mpsouterphysdist);
    }
    \draw[line width=\lw] (S) -- (S1.south);
    \draw[line width=\lw] (S1.east) to[out=-90,in=180] (S2.south);
    \draw[line width=\lw] (S2.east) to[out=-90,in=180] (S3.south);
\end{tikzpicture}
    }.
\end{equation}

\end{document}